\documentclass{jfm}
\usepackage{graphicx}
\usepackage{epstopdf, epsfig}

\usepackage{xcolor}

\shorttitle{Impact of a microfluidic jet onto a pendant droplet}
\shortauthor{M.A. Quetzeri-Santiago, I.W. Hunter, D. van der Meer and D. Fernandez Rivas}

\title{Impact of a microfluidic jet onto a pendant droplet}

\author{Miguel A. Quetzeri-Santiago\aff{1}  \corresp{\email{m.a.quetzerisantiago@utwente.nl and d.fernandezrivas@utwente.nl}},
  Ian W. Hunter\aff{2}, Devaraj van der Meer\aff{3}
 \and David Fernandez Rivas\aff{1,2}}

\affiliation{\aff{1}Mesoscale Chemical Systems Group, MESA+ Institute and Faculty of Science and Technology, University of Twente,
P.O. Box 217, 7500AE Enschede, The Netherlands
\aff{2}BioInstrumentation Laboratory, Department of Mechanical Engineering, Massachusetts Institute of Technology, Cambridge, Massachusetts 02139, USA
\aff{3}Physics of Fluids Group, Max-Planck Center for Complex Fluid Dynamics, MESA+ Institute, J.M. Burgers Center for Fluid Dynamics and Faculty of Science and Technology, University of Twente, P.O. Box 217, 7500AE Enschede, The Netherlands}

\begin{document}

\maketitle

\begin{abstract}
High speed microfluidic jets can be generated by a thermocavitation process: from the evaporation of the liquid inside a microfluidic channel, a rapidly expanding bubble is formed and generates a jet through a flow focusing effect. Here, we study the impact and traversing of such jets on a pendant liquid droplet. Upon impact, an expanding cavity is created, and, above a critical impact velocity, the jet traverses the entire droplet. We predict the critical traversing velocity (i) from a simple energy balance and (ii) by comparing the Young-Laplace and dynamic pressures in the cavity that is created during impact. We contrast the model predictions against experiments, in which we vary the liquid properties of the pendant droplet and find good agreement. In addition, we asses how surfactants and viscoelastic effects influence the critical impact velocity. Our results are relevant for the study of needle-free injections, where jets of similar velocities and dimensions are being used. Given the simplicity of our system we can systematically vary the target properties and unravel their effect on the impact dynamics, a true challenge when injecting into real skin.
\end{abstract}

\section{Introduction}

The impact of a solid or liquid object into a deep liquid pool generates a cavity with dynamics first described by \cite{worthington1908study}. Since then, research has focused on many topics, including the critical energy necessary for air entrainment into the pool, the collapse of the formed cavity, and the subsequent formation of Worthington jets~\citep{oguz1995air, lee1997cavity, zhu2000mechanism, lorenceau2004air, aristoff2009water, truscott2014water}. However, the projectiles studied usually have sizes in the range of 0.2 to 5 mm and an impact speed range of 1 to 10 m/s, and the pool is usually orders of magnitude larger than the projectile. In these cases, hydrostatic pressure has been found a major driver for the collapse and retraction of the cavity made on the liquid pool \citep{oguz1995air}. 
In contrast, in this paper we study (for the first time) the impact of micrometer-sized jets onto a compact liquid object, namely a droplets of $\sim 2$ mm. The jets travel at speeds of $\sim$ 20 m/s, and the hydrostatic pressure can be neglected due to the size of the created cavity.

\begin{figure}
\centering
\includegraphics[width=13cm]{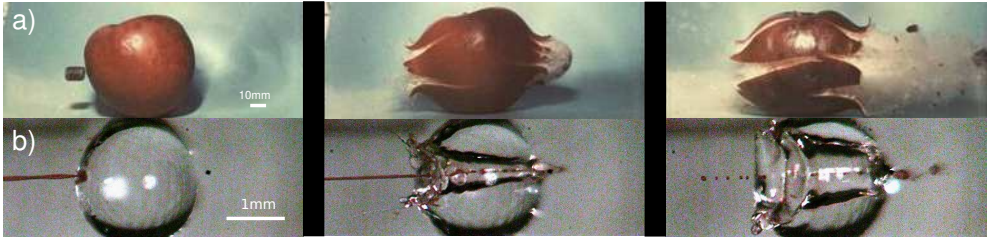}
\caption{Comparison between a sequence of images from, a) Harold E. Edgerton, Bullet through Apple, 1964, $U_{bullet} \sim 500$ m/s (reprinted with the permission of James W. Bales and Andrew Davidhazy).  The sequence is taken using the flash photography technique \citep{edgerton1931stroboscopic}, with a flash duration of $\approx$ 1/3 $\mu$s. The flash is triggered by an electronic circuit that reacts to the sound of the rifle shot. b) The impact of a liquid jet on a droplet; the video is recorded with a high-speed camera at 50k frames per second, the jet diameter is $D_{jet} =$ 100 $\mu$m and its impact velocity is $U_{jet} = 25.8$ m/s (see also movie 1 in the supplementary material). Apart from the striking aesthetic resemblance of the processes, the ratio between the projectile kinetic energy and the energy associated with the target's resistance to deformation is on the same order of magnitude, see Section \S\ref{ApplevsDrop}.}
\label{Fig1}
\end{figure}

Since, the dynamics of the aforementioned events take place in a few miliseconds, high-speed imaging is required to observe the phenomena. High-speed imaging was pioneered by scientists like Harold E. Edgerton, with his strobe-flash photography technique \citep{edgerton1931stroboscopic}. One of his most famous sequences of pictures is that of a 0.22 inch caliber bullet traversing an apple at $\approx 500$ m/s (figure \ref{Fig1}a). In this sequence, the apple is opaque, but, what if we could replace the target with a transparent object, i.e., a \textit{liquid apple}? This is precisely the case when we target a droplet with a liquid microjet. As shown in figure \ref{Fig1}, the aesthetic of a high-speed jet traversing a liquid droplet and Edgerton's pictures is strikingly similar. The difference being that,  with a translucent liquid, we can observe the impact dynamics inside the droplet, and besides producing impressive images, high-speed imaging facilitates the description of the fast phenomena, such as providing the projectile speed and the target deformation.

Our aim in this paper is to unravel the physics that govern the impact of liquid water jets on a pendant droplet of liquids with different properties. Ultimately, this knowledge targets at increasing our understanding of needle-free injection in skin. Needle-based injections are among the most common procedures in modern medicine, but pain and phobia affect patience compliance \citep{hamilton1995needle,sokolowski2010needle}. Needle-free injections with liquid jets have been proposed to solve some of these issues, with jet velocities as high as 100 m/s ---and volumes ranging from 10 nL to 1 mL \citep{prausnitz2004current, hogan2015needle, mohizin2018current}.

More recently, the potential of jets with 1-25 nL --micro-jets-- and superficial penetration in skin has been demonstrated~\citep{kiyama2019visualization, cu2019delivery, oyarte2020microfluidics, krizek2020needle}. However, for micro-jet injectors to become a reality, understanding the impact and penetration dynamics of the liquid into real tissue in real time is crucial~\citep{mercuri2021challenges}. Systematic studies aimed at quantifying the effect of skin properties during an injection are challenging, because the geometry, variability and opacity of the skin difficult optical access. In the current experiments, however, we can precisely vary and control the droplet properties and study the dynamics of jet penetration using droplets as skin surrogate.

The paper is structured as follows. The experimental procedure is outlined in \S\ref{ExperimentalSection}. In \S\ref{Predicting} we present two models to predict the critical jet velocity for traversing a droplet: In \S\ref{BlanceSection} we use an energy balance between the kinetic energy of the jet and the surface energy of the droplet. In \S\ref{Young-Laplace}, we employ the two-dimensional Rayleigh equation to obtain the cavity shape and combine it with the Young-Laplace equation to predict its collapse. Next, the model prediction is contrasted against experimental results in \S\ref{TraversingvsEmbedding}. Additionally, in \S\ref{CavityDynamics} and \S\ref{Observations} we present experimental results on the cavity advancing and retracting velocities with some observations on what occurs after the cavity collapse.

\section{Experimental Method}
\label{ExperimentalSection}

\begin{figure}
\centering
\includegraphics[width=13cm]{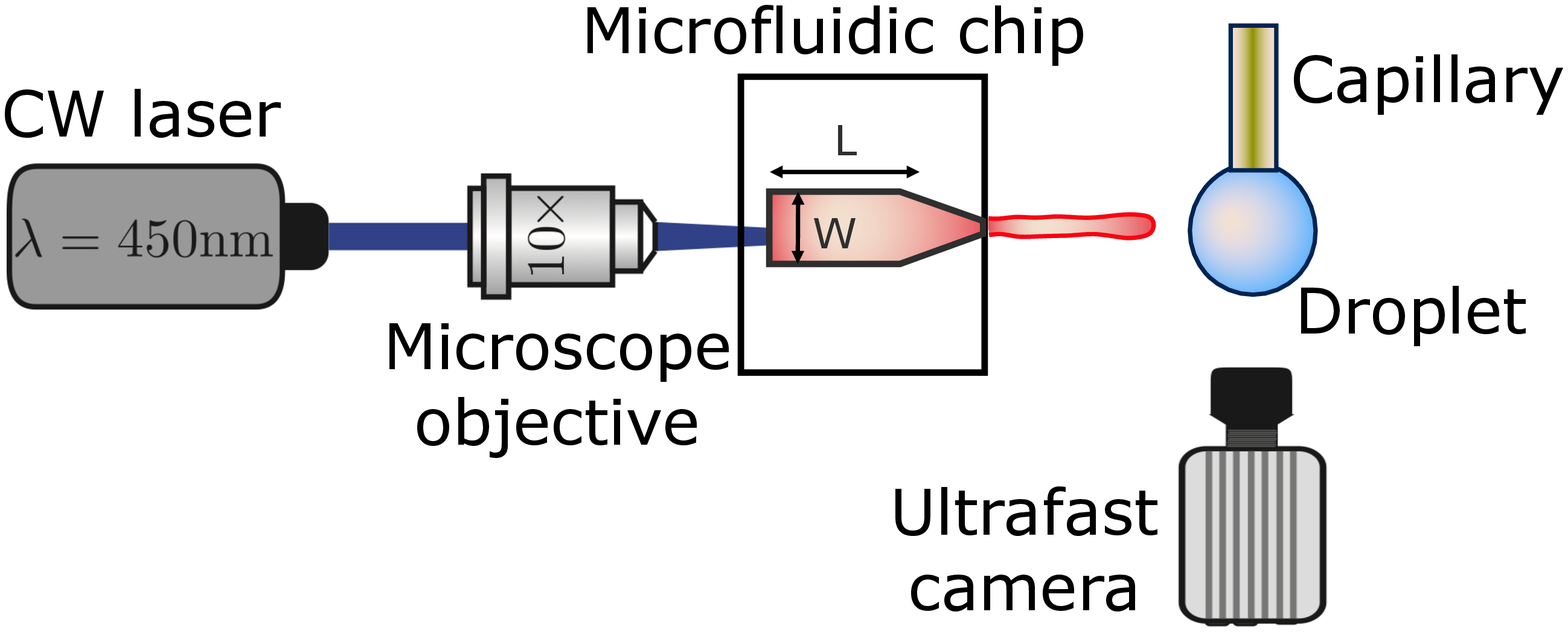}
\caption{Schematic of the experimental setup. Thermocavitation is obtained focusing a CW laser at the bottom of a microfluidic device with a microscope objective. The thermocavitated bubble expands and creates a liquid jet that is directed to a pendant droplet. The process is recorded with a high-speed camera with the illumination coming from a light source from the front so the cavity evolution can be observed.}
\label{Fig2}
\end{figure}

High-speed jets were generated from a thermocavitation process and directed to impact a pendant droplet of different liquids with varying properties. Thermocavitation refers to the phenomena where a liquid is vaporised locally by means of a focused laser, leading to bubble nucleation \citep{rastopov1991sound, padilla2014optic}. The expansion of the nucleated bubble can be controlled on a microfluidic chip to generate a jet through a flow-focusing effect \citep{rodriguez2017toward, oyarte2020microfluidics}. These jets may reach speeds in the order of 100 m/s, which, is enough to pierce the skin, and has potential for transdermal delivery of a liquid, particularly needle-free injections \citep{cu2019delivery, oyarte2019high}. However, in  the experiments described here, we restrict the jet velocity $U_{jet}$ to the range of 8 - 35 m/s, which is sufficient for a jet to traverse droplets of the liquids studied. Additionally, the diameter of the liquid jet was in the range of $D_{jet}$ = 50 - 120 $\mu$m. Both $U_{jet}$ and $D_{jet}$ were controlled by varying the laser spot size and power. 

The experimental setup is shown in figure \ref{Fig2}. A Borofloat glass microfluidic chip fabricated under cleanroom conditions is filled with a water solution containing a red dye (Direct Red 81, CAS No. 2610-11-9) at 0.5 wt. $\%$. The red dye enhances the laser energy absorption and facilitate the bubble nucleation. The microfluidic device has a tapered channel with an angle $\alpha = 15$ degrees to avoid swirling of the jet \citep{oyarte2020microfluidics}, nozzle diameter $d =$ 120 $\mu$m, channel length $L =1050$ $\mu$m and width $W = 600$ $\mu$m. The thermocavitation bubble is created by focusing a continuous wave laser diode (Roithner LaserTechnik, wavelength $\Lambda = 450$ nm and nominal power of 3.5 W), at the microchannel with a 10x microscope objective.  The liquids used were water, ethanol, aqueous solutions of glycerol, Triton x-100 and sodium-bis(2-ethylhexyl)sulfosuccinate (Aerosol OT) at different concentrations and polyethylene-oxide of varied molecular weight (PEO). Liquid droplets were created by holding a capillary tube with outer diameter of 360 $\mu$m, controlling the volume with a precision glass syringe and a syringe pump (Harvard PHD 22/2000). All chemical additives were bought from Sigma-Aldrich. The properties of the Newtonian and non-Newtonian liquids used are reported in table 1. The surface tension of all the liquids was measured with the Pendent Drop ImageJ plugin \citep{daerr2016pendent_drop}, and their shear viscosity with an Anton Paar MCR 502 rheometer. 

\begin{figure}
\centering
\includegraphics[width=13cm]{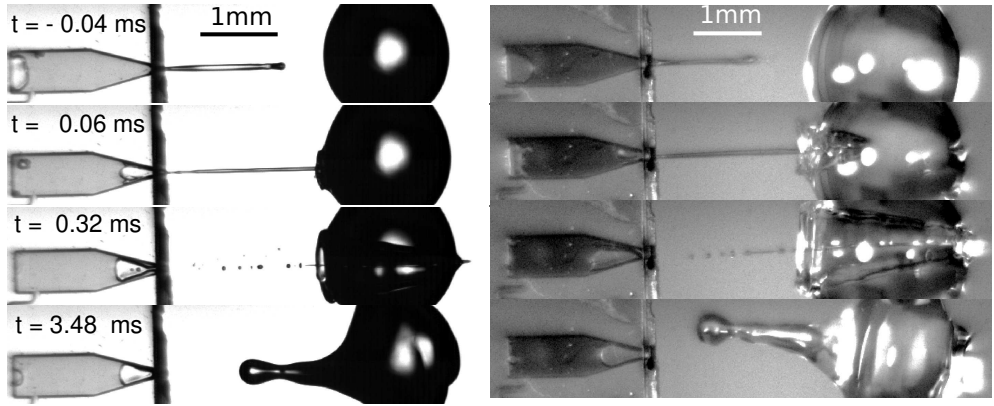}
\caption{Snapshot sequences of a jet impacting on a  water droplet with Triton X-100 at 3 CMC (left, movie 2 in the supplementary materials) and a pure water droplet (right, movie 3 in the supplementary materials). In the former $U_{jet} = 16.3$ m/s, $We_{jet} = 654$ and $D_{drop} =1.65$ mm. In the latter $U_{jet} = 17.1$ m/s, $We_{jet} = 325$ and $D_{drop} = 1.86$ mm. The imaging of the left sequence is done with a conventional shadowgraph illumination system. In contrast, the right sequence is taken with a front light illumination system, where the cavity dynamics can be more clearly observed. Times are taken relative to the impact moment at $t = 0$. The impact process is qualitatively the same for both of the droplets; a cavity is generated inside the droplet (t $\sim$ 0.06 ms), the jet traverses the droplet (t $\sim$ 0.032 ms) and a rebound Worthington jet is generated (t $\sim$ 3.48 ms). Time is taken from the impact moment t = 0 ms.}
\label{Fig3}
\end{figure}

The processes of bubble generation, jet ejection and impact on the liquid droplet were recorded with a Photron Fastcam SAX coupled with a 2x microscope objective. A typical experiment duration was $\sim 5$ ms and the camera resolution was set to $768 \times 328$ pixels$^2$ at a sample rate of $50$k frames per second with an exposure time of $2.5$ $\mu$s. Typical images obtained from the experiments are shown in figure \ref{Fig3}, where one observes how a water droplet is pierced by the liquid jet produced from the microchip on the left, using shadowgraph imaging (left) and direct lighting (right). Experiments were carried out with a typical shadowgraph configuration, and we switched to a front light illumination system to observe the cavity dynamics. In the front illumination system a white background was placed to enhance image contrast and increase the light reaching the camera sensor. Image analysis to extract the jet diameter, impact velocity and cavity dynamics was performed with a custom generated MATLAB script.  The shadowgraph imaging benefits from more light reaching the sensor, and thus a smaller camera exposure time can be used, leading to a better jet definition.  However, extracting information from the expanding cavity is impossible. In contrast, with front light imaging we can extract the information from the expanding cavity, but the jet is not as well defined as with shadowgraphy. 

\begin{table}
\small
  \caption{List of fluids used providing their shear viscosity $\mu$, surface tension $\gamma$ and density $\rho$. The viscoelastic relaxation time $\lambda$ is also shown for the polyethylene-oxide solutions.}
  \label{FluidPropertiesNewtonian}
  \begin{center}
  \begin{tabular*}{\textwidth}{@{\extracolsep{\fill}}lrrrc}
    \hline
   Fluid & $\mu$ (mPa s) & $\gamma$ (mN/m)& $\rho$ (kg/m$^3$) &  $\lambda$ (ms)\\
    \hline
    Ethanol & 1.04 & 26.3 & \ \ 789 & -\\
    Water & 1.0 & 72.1 & \ \ 998 & -\\
    Aqueous Glycerol 25 v$\%$ & 2.4 & 69.7 & 1071 & -\\
    Aqueous Glycerol 50 v$\%$ & 8.4 & 67.6 & 1142 & - \\
    Aqueous Glycerol 70 v$\%$ & 28.7 & 66.1 & 1193 & - \\
    Aqueous Glycerol 78 v$\%$ & 43.6 & 65.2 & 1212 & - \\
    Triton 0.2 CMC$\%$ & 1.0 & 43.9 & 998 & - \\
    Triton 1 CMC$\%$ & 1.0 & 30.8 & 998 & - \\
    Triton 3 CMC$\%$ & 1.0 & 32.5 & 998 & -\\
    Aerosol OT 1 wt.$\%$ (AOT 1$\%$) & 1.0 & 23.4 & 998 & - \\
    Aerosol OT 0.1 wt.$\%$ (AOT 0.1$\%$) & 1.0 & 24.1 & 998 & -\\
    Water $\&$ red dye 0.5 wt.$\%$ & 0.91 & 47.0 & 1000 & - \\
    PEO 100k 0.1 wt$\%$ & 1.03 & 63.2 & 996 & 0.006\\
    PEO 100k 1 wt$\%$ & 2.43 & 62.9 & 995 & 0.047\\
    PEO 100k 10 wt$\%$ & 50.8 & 62.5 & 1001 & 0.333\\
    PEO 600k 0.1 wt$\%$ & 1.56 & 63.1 & 996 & 0.307 \\
    PEO 600k 1 wt$\%$ & 21.7 & 62.9 & 998 &  1.317\\
  \end{tabular*}
  \end{center}
\end{table}

\section{Critical jet velocity}
\label{Predicting}

In this section, we  predict the critical velocity needed for a jet to traverse a droplet using two different approaches, (i)  by  using  a  simple  energy balance and (ii) by comparing the Young-Laplace and dynamic pressures in the cavity that is created during impact. In \S\ref{BlanceSection}, we start from an energy analysis of Edgerton's experiment of a bullet traversing an apple and subsequently transfer the argument to the droplet case of our current study. With this example we introduce the concept of kinetic energy of the projectile and the resistance of the target to being traversed. Moreover, we deduce the critical velocity of the jet by doing an energy balance between the kinetic energy of the jet and the surface energy of the droplet. Additionally, in \S\ref{Young-Laplace}, we use the Rayleigh's two dimensional equation to predict the shape of the cavity and predict its collapse with the Young-Laplace equation, thus finding the jet critical traversing velocity.  

\subsection{Energy balance between the jet kinetic energy and the droplet surface energy}
\label{BlanceSection}
\label{ApplevsDrop}

In his lecture titled \textit{How to Make Applesauce at MIT}, Edgerton presented his famous series of pictures of bullets traversing apples presented in figure \ref{Fig1}a. This set of images illustrated the traversing process, but did not reflect on the energy of the bullet or the energy of the apple. What would it take the apple to stop the bullet? Or equivalently, what would be the necessary speed for the bullet to get \textit{trapped} and \textit{embeded} inside the apple? 

In this section, we will answer these questions by using an energy balance between the kinetic energy of a bullet $E_{k_{bullet}} = M_{bullet}U_{bullet}^2/2 $, where $M_{bullet}$ is the mass of the bullet, and the toughness of an apple $T_{apple}$, which we define as its ability to absorb energy by elastoplastic deformation without fracturing. Hence, by doing the energy balance, the critical velocity for the bullet to penetrate the apple may be written as

\begin{equation}
    U_{bullet}^* = \sqrt{2\,T_{apple}/M_{bullet}}.
\end{equation}

The mass of a .22 caliber bullet is $M_{bullet} \sim 10$ g and the apple toughness is  $T_{apple} \sim 10$ J \citep{grotte2001mechanical}. Therefore, $U_{bullet}^* \sim 45$ m/s, which is at least one order of magnitude smaller than the typical velocities reached by .22 caliber bullets, $U_{bullet} \sim 500$ m/s. Consequently, it is understandable that the apple is traversed by the bullet in Edgerton's photographs.

For our liquid jet, the kinetic energy is $E_{k_{jet}} \sim (\pi/8) \rho_{jet}U_{jet}^2 D_{jet}^2 H_{jet}$, with $\rho_{jet}$ and $H_{jet}$ the density and length of the jet respectively, and the resisting force of the droplet is dominated by its surface energy. For the critical conditions where the jet traverses the droplet, the jet kinetic energy transforms into the surface energy of the cavity generated at impact. For simplicity, assuming that the cavity geometry is cylindrical, the cavity surface energy is $E_{\gamma_c} \sim \pi D_c D_{drop} \gamma_{drop}$, with $\gamma_{drop}$ the droplet surface tension and $D_c$ the cavity diameter. Here, $D_c$ is constrained by $D_{drop}$ and as shown in figures \ref{Fig3} and \ref{Fig4}, $D_c \sim D_{drop}$. Also, since the velocity of the tip of the cavity is approximately half the jet velocity $U_c \sim \frac{1}{2} U_{jet}$, the total length of the jet would not contribute to the traversing process but only a part of it, namely $H^*_{jet} \approx 2D_{droplet}$ \citep{bouwhuis2016impact}. Using this limiting value $H^*_{jet}$, the jet critical velocity for droplet traversing is

\begin{equation}
    U_{jet}^{\dagger} \sim \left(\frac{4 \gamma_{drop} D_{drop}}{\rho_{jet} D_{jet}^2}\right)^{1/2}.
    \label{UCriticalBalance}
\end{equation}

Defining the relevant Weber number of the jet as $We_{jet} = \rho_{jet} U_{jet}^2 D_{jet}/\gamma_{drop}$, and substituting Eq. \ref{UCriticalBalance} we find the critical minimal Weber number needed to traverse the droplet

\begin{equation}
    We_{jet}^{\dagger} = \frac{4 D_{drop}}{D_{jet}}.
    \label{WeCriticalBalance}
\end{equation}

Substituting typical values of a jet impacting a water droplet in our experiments ($\rho_{jet} \sim 1000$ kg/m$^3$, $D_{jet} \sim 100$ $\mu$m $D_{drop} \sim$ 2 mm and $\gamma_{drop} \sim$ 0.07 mN/m), we obtain $U^{\dagger}_{jet} \sim 7.5$ m/s or $We^{\dagger} \sim 80$.

Now, we have all the ingredients to do a scaling comparison between a bullet traversing an apple and a jet traversing a droplet. Taking the values of $U_{jet}$ and $U_{bullet}$ from the experiments in figure \ref{Fig1}a) and \ref{Fig1}b (which are well above the critical value for penetration in both cases) and the target toughness (toughness for an apple and surface energy for a droplet), we get that $E_{k_{bullet}}/T_{apple} \sim E_{k_{jet}}/E_{\gamma_{c}} \sim 100$. Therefore, the relative energies involved in both processes are of the same order of magnitude, indicating that the traversing phenomena in both cases share more than aesthetic similarities. Nevertheless, after impact, the fractured apple does not possess the restoring force a liquid droplet has, namely, the surface tension. This is the cause of the much appreciated fact that we did not have to deal with substantial amounts of debris after our experiments.

\subsection{Comparison between the Young-Laplace and dynamic pressures of the cavity}
\label{Young-Laplace}

\begin{figure}
\centering
\includegraphics[width=8cm]{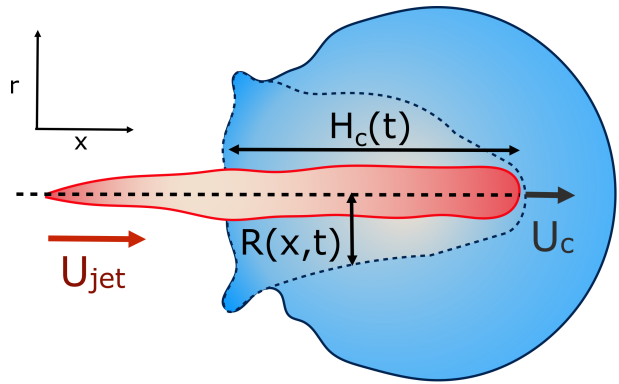}
\caption{Diagram illustrating the parameters used in this section. The jet impacts the pendant droplet from the left with velocity $U_{jet}$, creating a cavity with increasing depth in time $H_c(t)$, where the velocity of the apex of the cavity is indicated by $U_c$ ($= \dot{H}_c$) and whose  radius depends on time and the position $R(x,t)$. Here, $x$ is the direction along which the jet travels and $r$ is its perpendicular radial direction.}
\label{Fig4}
\end{figure}

Considering the mass of a cylindrical liquid jet with radius $R_{jet}$ and length $H_{jet}$ falling into a pool of the same liquid, air is entrained in the pool at sufficiently energetic impacts, i.e., $We >> 1$ and $Re >> 1$ \citep{oguz1995air}. Additionally, the cavity dynamics and the air entrainment depend on the aspect ratio of the jet. The limiting cases are $H_{jet}/R_{jet} \to \infty$, corresponding to the impact of a continuous jet, and $H_{jet}/R_{jet} \to 1$, where the case of a droplet impact into a liquid pool is recovered \citep{oguz1995air, kiger2012air}. For the former case, the apex of the cavity advances with a velocity $U_c = \frac{1}{2} U_{jet}$, therefore, the depth of the cavity can be estimated as $H_c = \frac{1}{2}U_{jet} t$ \citep{oguz1995air, bouwhuis2016impact, speirs2018water}. 

In the cavity formation of a droplet impacting a liquid surface, the process is mainly inertial during the first instants, with surface tension becoming important at the moment near the maximum depth of the cavity $H_{max}$ \citep{bouwhuis2016impact}. Additionally, \cite{deng2007role} showed that viscous dissipation accounts for $\sim 1.4~\%$ of the initial kinetic energy loss of a water droplet of $D = 2.5$ mm impacting a liquid pool. Therefore, assuming that the cavity shape is slender and the process is inertia dominated, i.e., neglecting viscous dissipation, we can apply the two-dimensional Rayleigh equation in cylindrical coordinates to predict the cavity shape \citep{bergmann2006giant, eggers2007theory},

\begin{equation}
    \left(R \frac{d^2R}{dt^2} + \left(\frac{dR}{dt}\right)^2\right)\log \frac{R}{R_{\infty}} + \frac{1}{2}\left(\frac{dR}{dt}\right)^2 \approx \frac{ \gamma}{\rho R},
\end{equation}

where $R(x,t)$, is the radius of the cavity and $x$ is the position of the cavity on the horizontal direction and $R_{\infty}$ is an external length scale (see figure \ref{Fig4}). Following the argument of \cite{bouwhuis2016impact}, during the first instants of the cavity formation, inertia dominates and the dynamics are determined by $R(d^2R/dt^2) + (dR/dt)^2 \approx 0$. Solving this equation we get that $R(t) \sim (t - t_0)^{1/2}$, where $t = H_c/U_c$ and $t_0 = x/H_c$, and the approximate cavity profile is \citep{bouwhuis2016impact},

\begin{equation}
R(x,t) \sim \sqrt{\frac{D_{jet}}{2} (H_c - x)} = \sqrt{\frac{D_{jet}}{2} (U_{c}t - x)}.
\label{CavityProfile}
\end{equation}

The time $t_c$ at which surface tension can influence the cavity walls can be predicted by comparing the dynamic pressure of the radially expanding cavity and the Young-Laplace pressure based on the azimuthal curvature of the cavity,

\begin{equation}
    \rho \left( \frac{d R_{x = 0}}{dt} \right)^2 \sim \frac{2 \gamma}{R_{x=0}},
    \label{Laplace}
\end{equation}

where $R_{x=0} = R(0, t)$ is the cavity radius at the jet impact point $x = 0$. Taking the cavity profile from Eq. \ref{CavityProfile}, we get $R_{x = 0}(t) = \sqrt{1/2}\sqrt{D_{jet} U_c t}$, and $d R_{x = 0}/dt \sim D_{jet} U_c/2^{3/2} \sqrt{D_{jet} U_c t}$ \citep{bouwhuis2016impact}. Therefore,

\begin{equation}
    t_c \sim \frac{\rho_{jet}^2 D_{jet}^3 U_c^3}{128 \gamma_{drop}^2},
    \label{t_c}
\end{equation}

\noindent and

\begin{equation}
    H_{max} \sim U_c t_c.
    \label{H_max}
\end{equation}

The condition for the jet to traverse the droplet is that $D_{drop} < H_{max}$. Using that $U_c \sim  \frac{1}{2}U_{jet}$ and Eqs. \ref{t_c} and \ref{H_max}, the critical impact velocity for the jet to traverse the droplet is

\begin{equation}
    U_{jet}^* \sim 8 \left(\frac{\gamma_{drop}^2 D_{drop}}{2\rho_{jet}^2 D_{jet}^3}\right)^{1/4},
    \label{UCriticalLaplace}
\end{equation}

\noindent and 

\begin{equation}
    We_{jet}^* \sim 64\left(\frac{D_{drop}}{2 D_{jet}}\right)^{1/2}.
    \label{WeCriticalLaplace}
\end{equation}

For a jet impacting a water droplet, $D_{drop} = 2$ mm, $\gamma_{drop} = 0.072$ N/m, $D_{jet} = 100$ $\mu$m and $\rho_{jet} = 1000$ kg/m$^3$, we obtain that the critical velocity needed to traverse the droplet is $U_{jet}^* \approx 12$ m/s, which is $\sim 40 \%$ larger than $U_{jet}^*$ obtained from equation (\ref{UCriticalBalance}). Similarly, $We^* \approx 200$ which is about twice as large as for equation (\ref{WeCriticalBalance}). 

While the results in equations (\ref{WeCriticalBalance}) and (\ref{WeCriticalLaplace}) are of the same order of magnitude, their dependence on the ratio $D_{drop}/D_{jet}$ is different, namely linear in equation (\ref{WeCriticalBalance}) whereas in equation (\ref{WeCriticalLaplace}), there is a square root dependence. This discrepancy arises from the difference in the geometric shape of the generated cavity that was assumed in the two approaches, resulting in a different surface energy. Indeed, a very simple cylindrical geometry was assumed during the energy balance method. In contrast, deriving equation (\ref{WeCriticalLaplace}) using the Rayleigh equation, leads to a more rigorous description of the cavity shape. Therefore, we consider the latter model to be more accurate and in the following section will compare our experimental data to equation (\ref{WeCriticalLaplace}).

\section{Results and discussion}

In this section we will describe our experiments on the traversing of the jet through the droplet and compare them to the above criterion. Furthermore, we modify the criterion to include the concept of dynamic surface tension of the droplet $\gamma_{dyn}$ in the case of surfactant covered droplets. After that we will briefly discuss the cavity dynamics, focusing on the motion of the apex of the cavity inside the droplet. Finally, we comment on our observations for droplets containing surfactants and non-Newtonian liquids.

\subsection{Critical velocity for traversing}
\label{TraversingvsEmbedding}

\begin{figure}
\centering
\includegraphics[width=13cm]{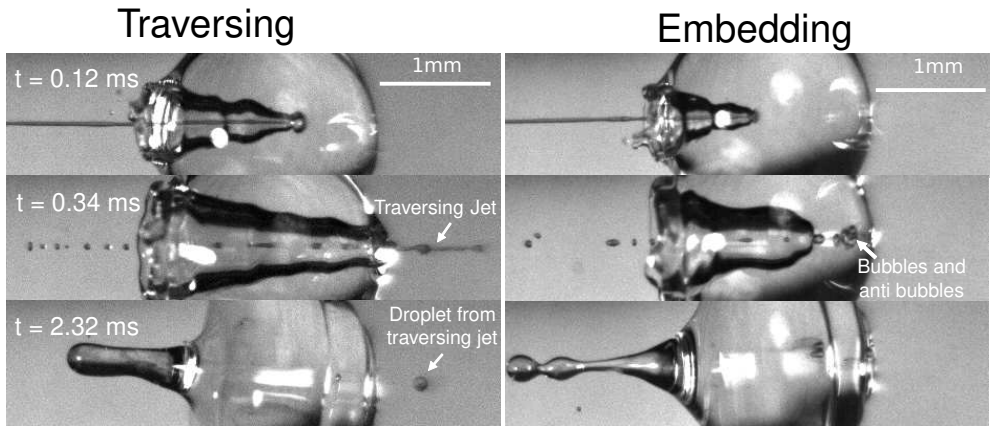}
\caption{Left, snapshots of a liquid jet impacting ($We_{jet} = 587$) and traversing a PEO 100k 0.1wt $\%$ droplet ($D_{drop} = 2.09$ mm). Here, we observe that the jet continues its trajectory even after going through all the droplet. Right, snapshots of a liquid jet impacting ($We_{jet} = 362$) and getting embedded in a PEO 100k 0.1wt $\%$ droplet ($D_{drop} = 2.08$ mm). During embedding, bubbles and antibubbles can be created, see movie 4 in the supplementary materials. We note that in  both image sequences a rebounding Worthington jet is observed at t =  2.32  ms}
\label{Fig5}
\end{figure}

We start the discussion of our experimental results by making a qualitative description of the observed phenomena. Figure \ref{Fig5} shows an image sequence from two typical experiments. Upon impact of the jet onto the droplet, a cavity is generated inside the droplet. The cavity diameter and depth increase with time and its growth rate is dependent on the impact conditions \citep{speirs2018water}. At a velocity above a critical value, the jet traverses the droplet completely, as is observed in the left panel of figure \ref{Fig5}. In contrast, if the jet velocity is not sufficiently large, the jet gets \textit{embedded} in the droplet and bubbles and anti-bubbles may be created, as in the right panel of \ref{Fig5} (see also \cite{song2020criteria}). Finally, and irrespective of which of these two scenarios applies a rebound Worthington jet is generated.

Now we move on to verifying the validity of the traversing criterion expressed in the critical Weber number obtained in equation (\ref{WeCriticalLaplace}), for varying droplet properties. To compare the experimental data and the model presented in section \S\ref{Young-Laplace}, we use the ratio between the experimentally obtained Weber number $We_{jet}$ and the expected critical Weber number $We_{jet}^*$ from equation (\ref{WeCriticalLaplace}). Additionally, for the droplets that contain surfactants we need to take into account that, when the jet impacts the droplet and the cavity starts to form, new surface area is created and the surface density of the surfactant decreases. Therefore, the surface tension locally increases from the surface tension measured at equilibrium and the cavity presents a \textit{dynamic} surface tension $\gamma_{dyn}$ \citep{speirs2018entry}. Consequently, in the surfactant case we re-define the Weber number as $We_{dyn} = \frac{\rho_{jet}U_{jet}^2D_{jet}}{\gamma_{dyn}}$, i.e., using $\gamma_{dyn}$ in its definition, and divide it by the critical value $We^*_{jet}$ leading to, 

\begin{equation}
    \frac{We_{dyn}}{We^*_{jet}} = \frac{We_{jet}}{We_{jet}^*}\left(\frac{\gamma_{drop}}{\gamma_{dyn}}\right) = \frac{\rho_{jet} U_{jet}^2 D_{jet}^{3/2}}{(2^{11/2}) \gamma_{dyn} D_{drop}^{1/2}}.
    \label{GammaDyn}
\end{equation}

Clearly in the above equation, for the droplets that do not contain sufactants (the glycerol solutions, water and ethanol) we just insert $\gamma_{dyn} = \gamma_{drop}$. For Triton X-100 solutions, the dynamic surface tension can be assessed by the diffusion scale $T_D$, which is the time for the surface tension to decrease from the surface tension of water to the equilibrium surface tension \citep{bobylev2019influence}. The diffusion scale depends on the diffusion coefficient of the surfactant (for Triton X-100 $\delta= 2.6\times 10^{-10}$ m$^2$/s), the maximum surface concentration of surfactant ($\Gamma = 2.9 \times 10^{-6}$ mol/m$^2$), the Langmuir equilibrium adsorption constant ($K = 1.5 \times 10^3$ m$^3$/mol) and its volume concentration $C$ \citep{bobylev2019influence, ferri2000surfactants}. For the 3 CMC Triton X-100 solution (the largest concentration used in these experiments), $T_D \sim 70$ ms, while the characteristic timescale of the traversing/embedding process is $\sim 0.5$ ms, i.e. two orders of magnitude smaller. Hence, the dynamic surface tension does not have enough time to reach the measured equilibrium surface tension. Therefore, we do not expect the equilibrium surface tension of Triton X-100 solutions to be relevant in the jet traversing process. Consequently, we assume the dynamic surface tension $\gamma_{dyn}$ of the Triton X-100 solutions to be that of water.

In contrast, AOT being a vesicle surfactant can migrate faster than micelle surfactants such as Triton X-100 \citep{song2017controlling, wang2019wettability}. In addition, it was shown that at $\sim 10$ ms the dynamic surface tension of an AOT solution at 1 wt. $\%$ can decrease to a value of $\sim 32$ mN/m \citep{song2017controlling}. Therefore, we assume that $\gamma_{dyn}$ for the AOT solutions is $\sim 32$ mN/m. We should note however, that AOT dynamics are more complex than those of Triton, and characterisation using a single time scale is an oversimplification. 

Figure \ref{Fig5} shows the experimental results of traversing and embedding impact cases as a phase diagram, where on the vertical axis we plot the ratio of the (dynamic) Weber number $We_{dyn}$ and the expected critical Weber number $We_{jet}^*$, using equation (\ref{GammaDyn}) such that based upon the model described in \S\ref{Young-Laplace} we would expect a transition at $We_{dyn}/We_{jet}^* = 1$. On the horizontal axis we separate the liquid properties of the droplet by plotting the Ohnesorge number, defined as $Oh_{drop} = \mu_{drop} /\sqrt{\rho_{drop} \gamma_{drop}}$, which is the ratio between viscous forces to inertial and surface tension forces, and has the advantage that is a material property, i.e. it is independent of the dynamics. Open symbols in figure \ref{Fig5} represent cases where the jet was observed to traverse the droplet, (as in figure \ref{Fig5} left) and solid symbols represent the situation where the jet does not traverse the droplet, i.e., becomes embedded as seen in figure \ref{Fig5} right.

\begin{figure}
\centering
\includegraphics[width=13cm]{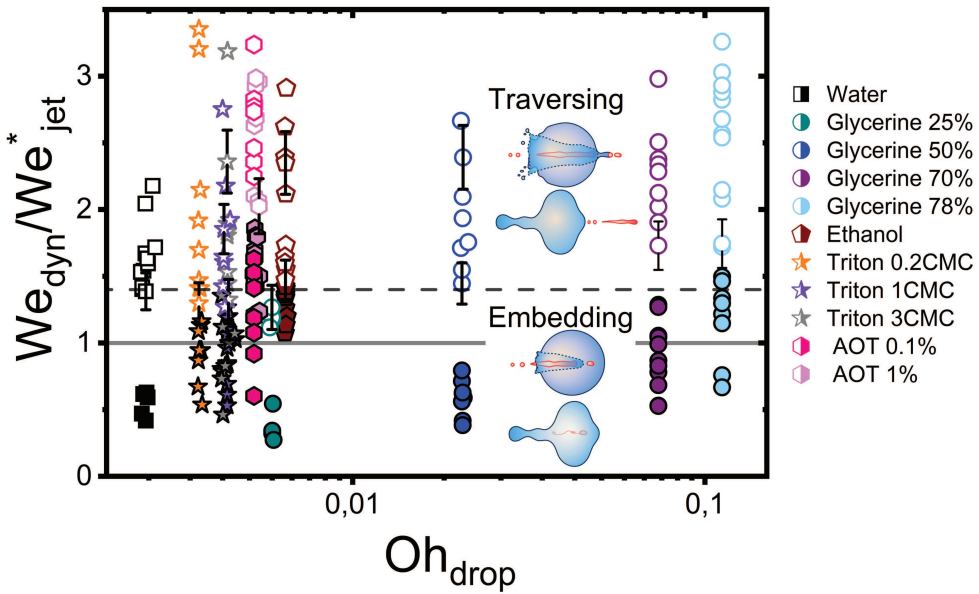}
\caption{Phase diagram with the rescaled Weber number $We_{dyn}/We^*_{jet}$ on the vertical axis and the Ohnesorge number $Oh_{drop}$ on the horizontal axis. Open symbols represent cases where the jet traverses the droplet, while solid ones stand for the embedding case. The grey line corresponds to $We_{dyn}/We^*_{jet} = 1$ and the dashed line is $We_{dyn}/We^*_{jet} = 1.4$ obtained by averaging the minimum value of observed traversing for all the liquids (excluding the AOT solutions). The experimental data is in good agreement with the model, i.e. for each liquid, most of the open symbols lie above the dashed line and conversely for closed symbols. Uncertainty was calculated for all the experimental data and example error bars are shown at selected points, where the uncertainty was found to increase linearly with the rescaled Weber number.}
\label{Fig6}
\end{figure}

From figure \ref{Fig6}, we observe that most of the open symbols lie above the same approximate value $\sim 1.4$ and conversely for closed symbols. The exception is formed by the data for the AOT solutions, where it is possible that $\gamma_{dyn}$ is underestimated as $\gamma_{dyn}\approx 32.2$ mN/m, and in fact lies closer to the surface tension of water $72.1$ mN/m. An accurate measurement of the dynamic surface tension in such timescales is challenging \citep{speirs2018entry,alvarez2010microtensiometer} and is out of the scope of this work. However, as demonstrated by our results, the dynamic surface tension can play a pronounced role for different dynamic conditions. Therefore, we can safely conclude that the impact process is initially dominated by inertia and that surface tension is the major opposing force.

\begin{figure}
\centering
\includegraphics[width=10cm]{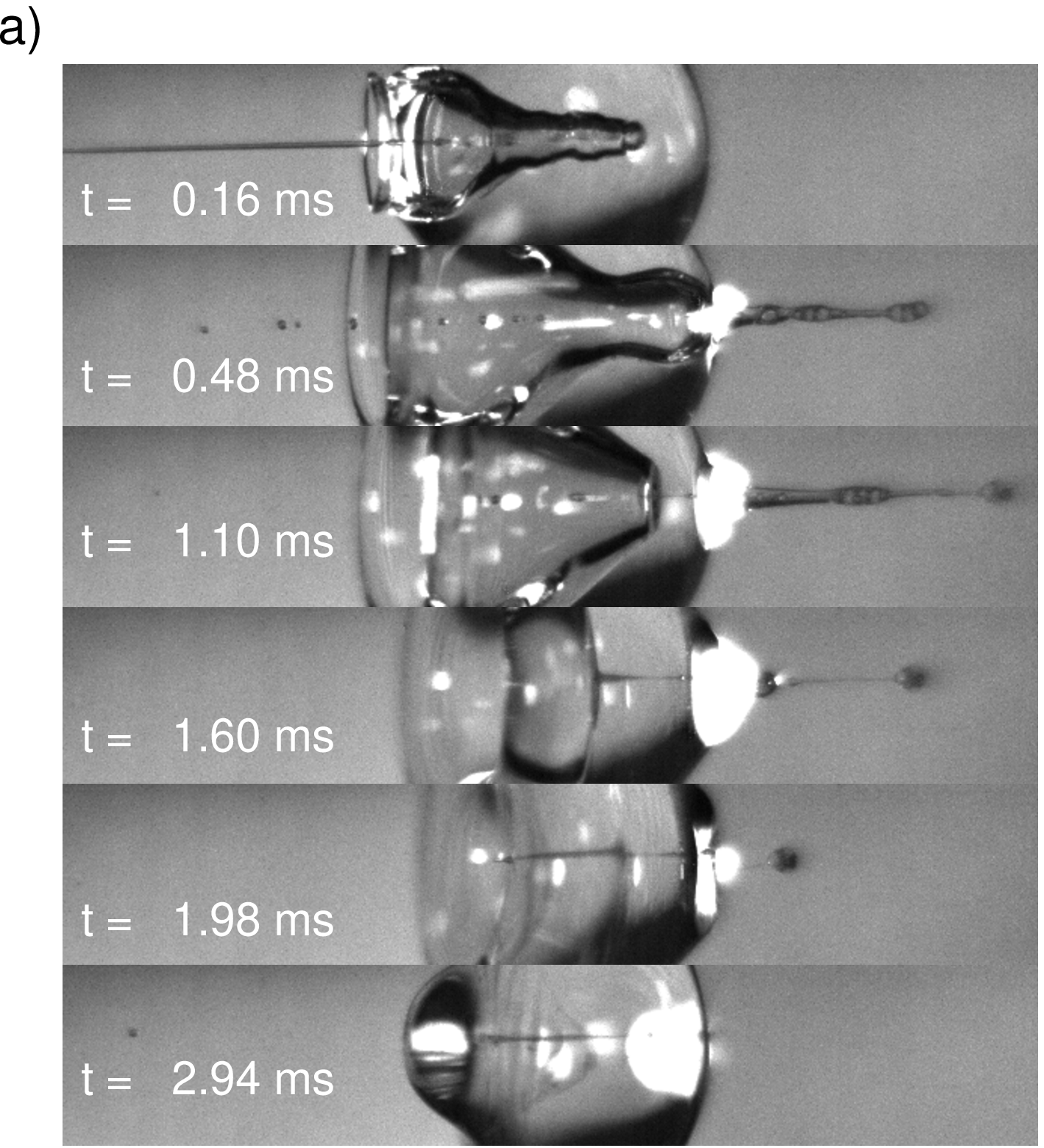}\vspace{0.2cm}
\includegraphics[width=10cm]{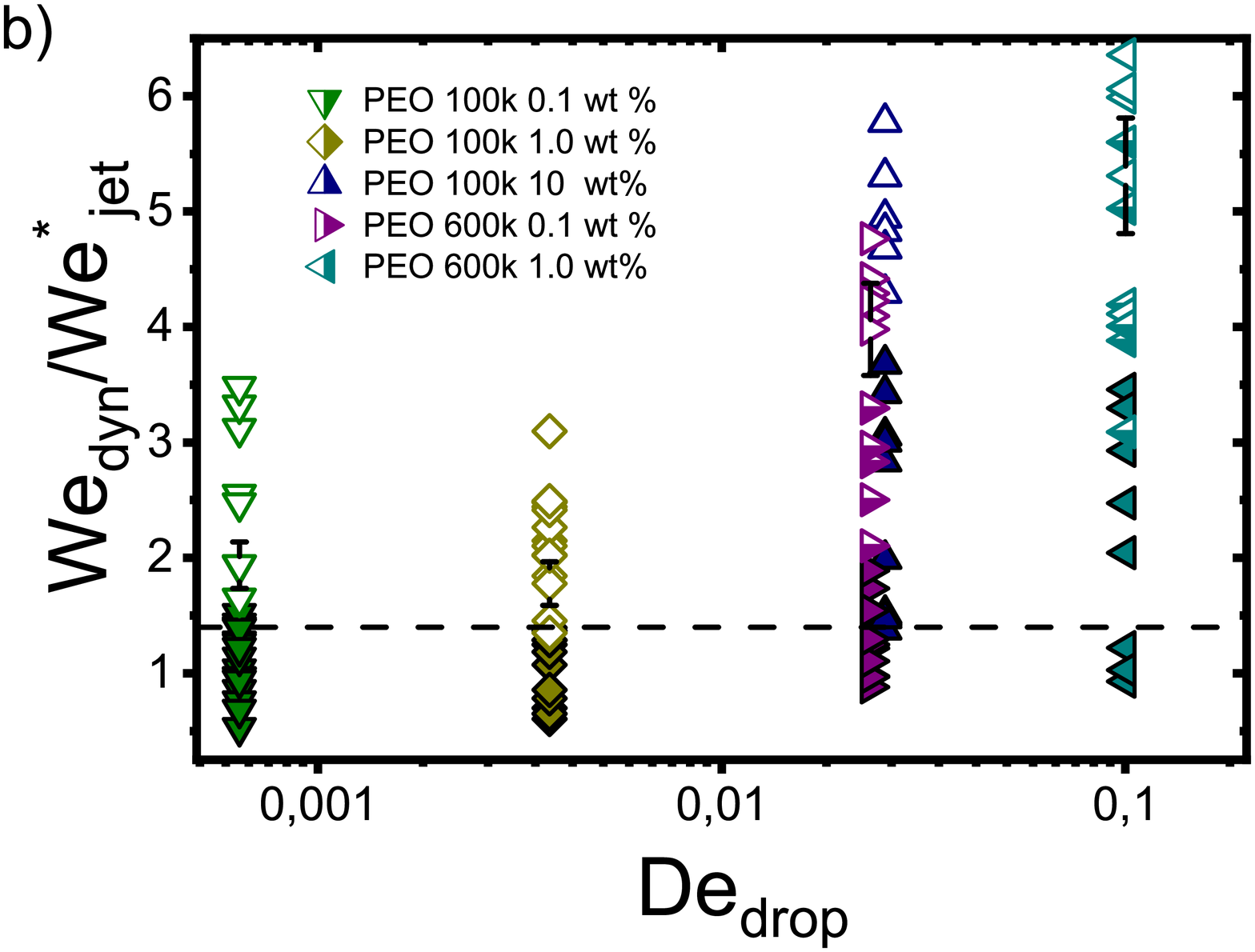}
\caption{a) Snapshots of a liquid jet impacting a water droplet with PEO 600k 1 wt$\%$ showing pearling (see movie 5 in the supplementary materials). A jet with  $U_{jet} = 33.5$ m/s and $We_{jet} = 1160$ impacts a droplet with $D_{drop} = 2.21$ mm. The jet travels a distance larger than $D_{drop}$, but due to the viscoelastic effects of the droplet, it gets sucked back into the droplet. b) Phase diagram in terms of $De$ and $We^*_{dyn}$. Open symbols represent cases where the jet traverses the droplet, solid ones stand for the embedding case and half filled ones represent pearling. The dashed line represents the transition from traversing to embedding cases observed for Newtonian liquids. Uncertainty was calculated for all the experimental data and example error bars are shown at selected points, where the uncertainty was found to increase linearly with $We_{dyn}/We^*_{jet}$.}
\label{Fig7}
\end{figure}

Turning to the viscoelastic droplets, figure \ref{Fig7}b shows data for the jet traversing and embedding impact cases for droplets consisting of the PEO solutions. In this figure, we plot $We_{dyn}/We^*_{jet}$ against the Deborah number defined as $De = \lambda/ \tau_c$, where $\lambda$ is the relaxation time of the polymer (see table 1) and $\tau_c = (\rho_{drop} D_{drop}^3/\gamma_{drop})^{1/2}$ is the capillary timescale. We use this definition of the Deborah number to our PEO solution droplets, as we expect to observe deviations from the Newtonian behaviour when $\lambda$ becomes comparable to the scale at which surface tension starts to influence the cavity dynamics, i.e. at the capillary time scale $\tau_c$. In figure \ref{Fig7}, open and closed symbols again represent traversing and embedding cases respectively and half-filled symbols denote an intermediate state between traversing and embedding, which we call \textit{pearling}. During pearling, the jet travels a distance larger than $D_{drop}$ after impact and thus protrudes from the droplet, but due to the viscoelastic properties of the liquid gets sucked back into the droplet, as visualised in the experimental snapshots of figure \ref{Fig7}a. 

From figure \ref{Fig7}b that the traversing and embedding process for the PEO solutions with $De \lesssim 4 \times 10^{-3}$, is similar for Newtonian liquids, leading to the same threshold value $We_{jet}^*$, showing that the viscoelastic effects are weak. However, as $De$ increases, i.e., when the viscoelastic timescales become increasingly comparable to the capillary time, the jet needs larger speeds to traverse the droplet. This is in line with previous experiments where by increasing the elastic modulus of gelatin the cavity depth of an impacting sphere would decrease, keeping the impact velocity constant \citep{kiyama2019gelatine}. These results show that viscoelastic properties as described by $De$ significantly change the traversing dynamics. This is crucial information when trying to understand needle-free injections on skin, as it had been shown that skin has viscoelastic properties \citep{fung2013biomechanics}. However, conducting systematic studies trying to quantify the influence of skin properties during injection processes is challenging, because of high variability from person to person and even between different parts of the body \citep{fung2013biomechanics,joodaki2018skin}. Furthermore, studying the viscoelastic properties of skin is in itself challenging given the opacity of skin \citep{crichton2011viscoelastic, graham2019stiff}. In this context, our results present information about the characteristics of the impact with a simpler system than skin, isolating the effects of individual  material properties of the target from the enormous complexity of skin.

\subsection{Cavity dynamics}
\label{CavityDynamics}

\begin{figure}
\centering
\includegraphics[width=6.5cm]{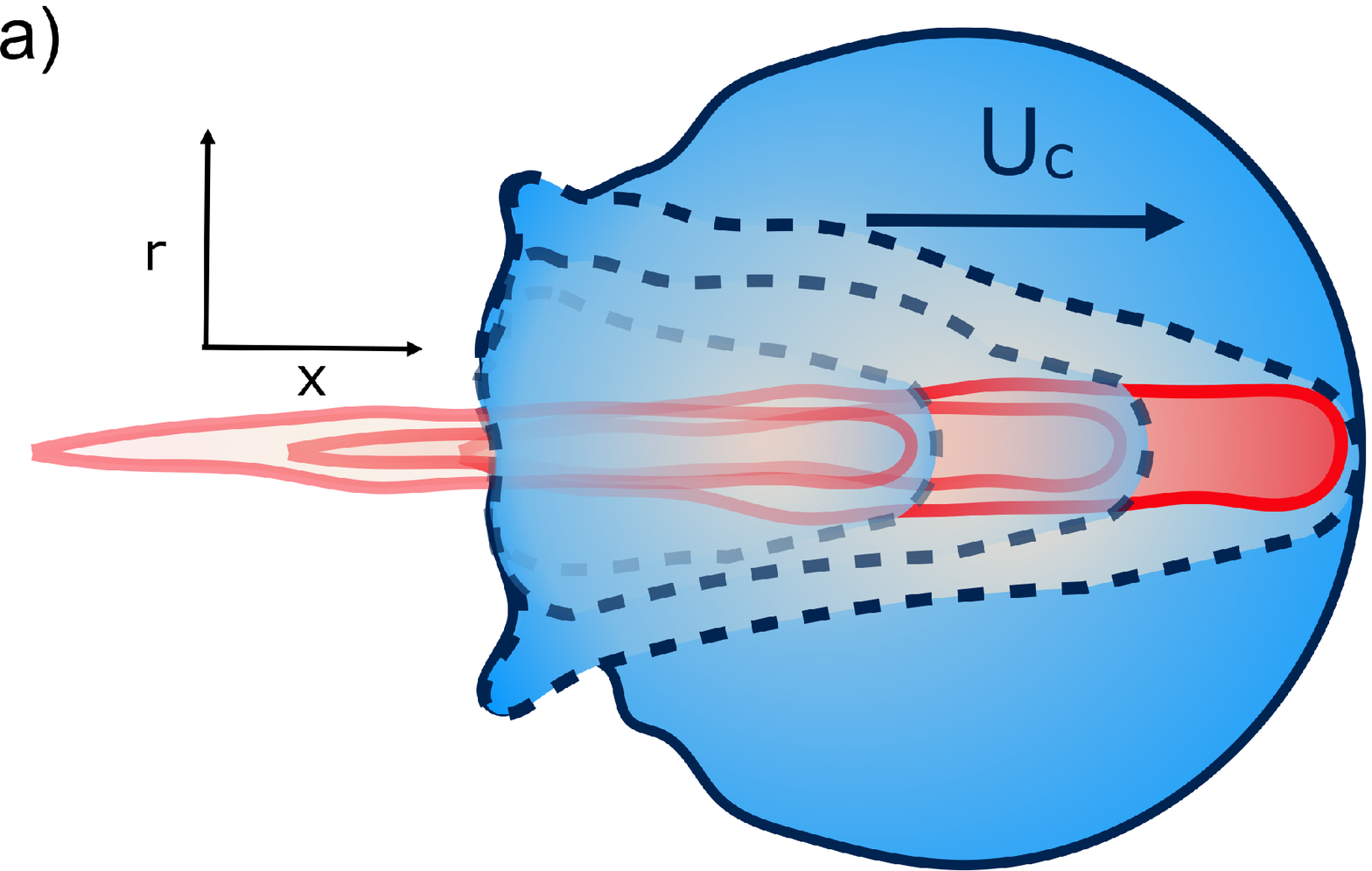}
\includegraphics[width=6.5cm]{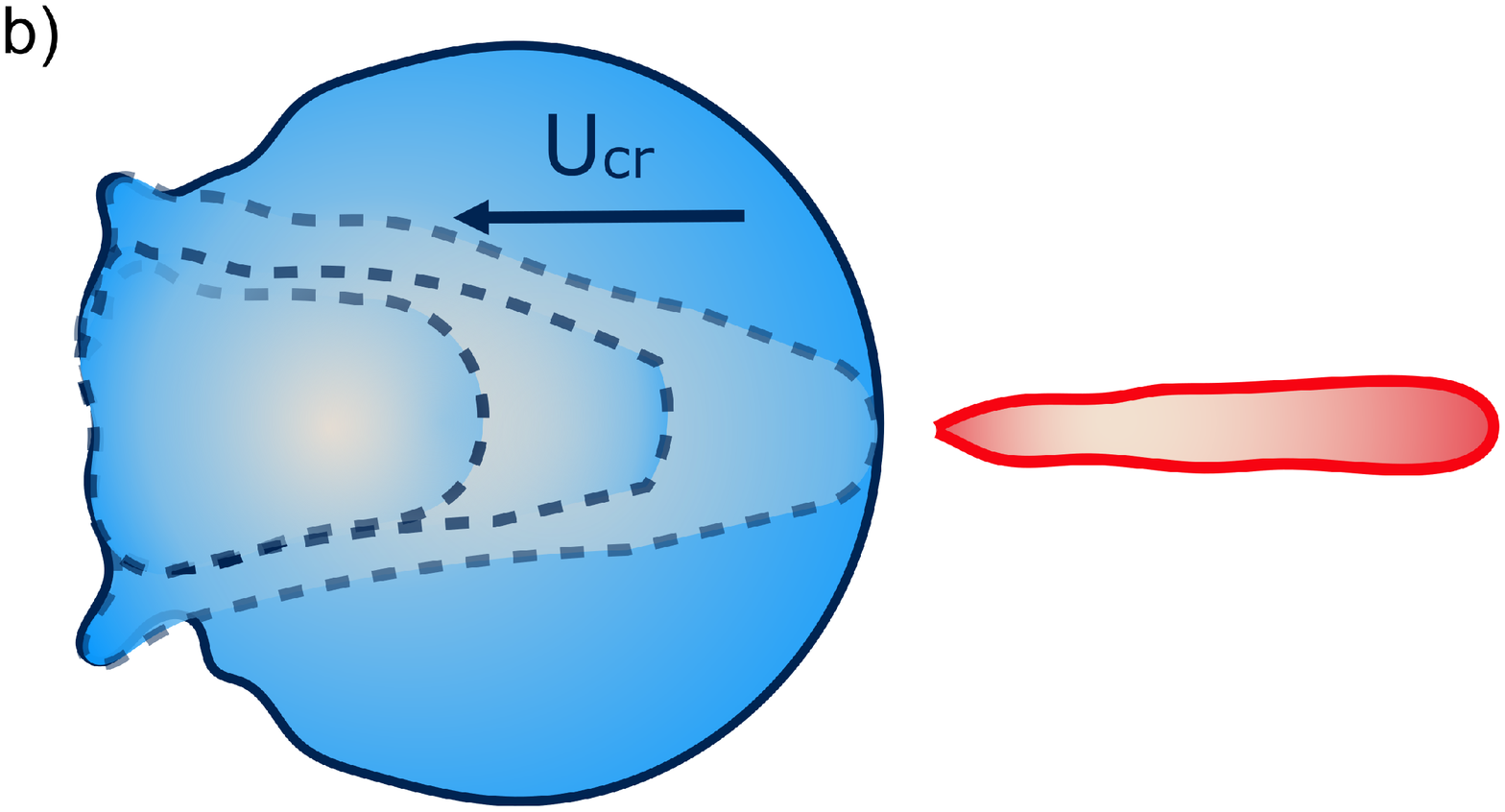}
\includegraphics[width=6.5cm]{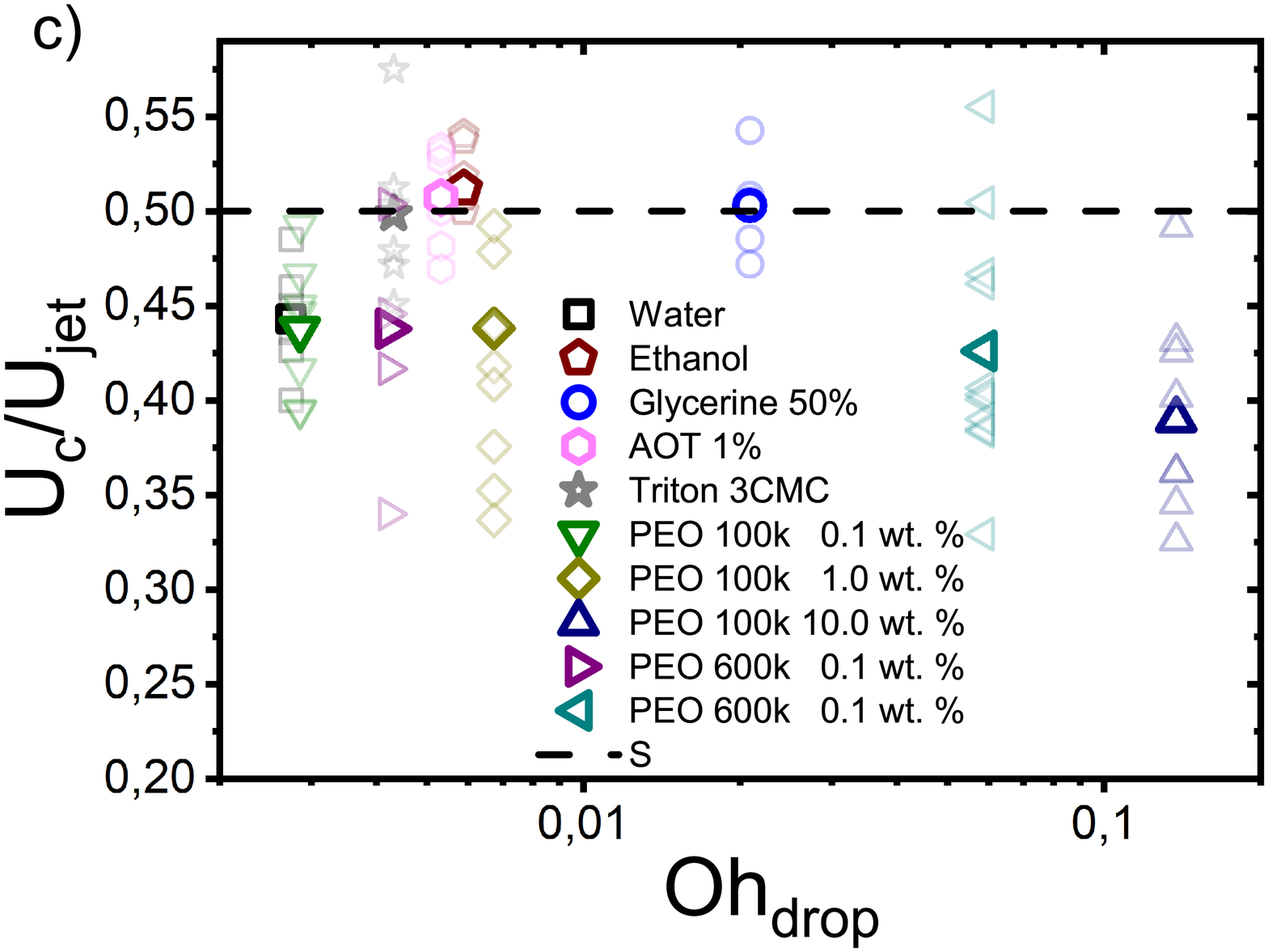}
\includegraphics[width=6.5cm]{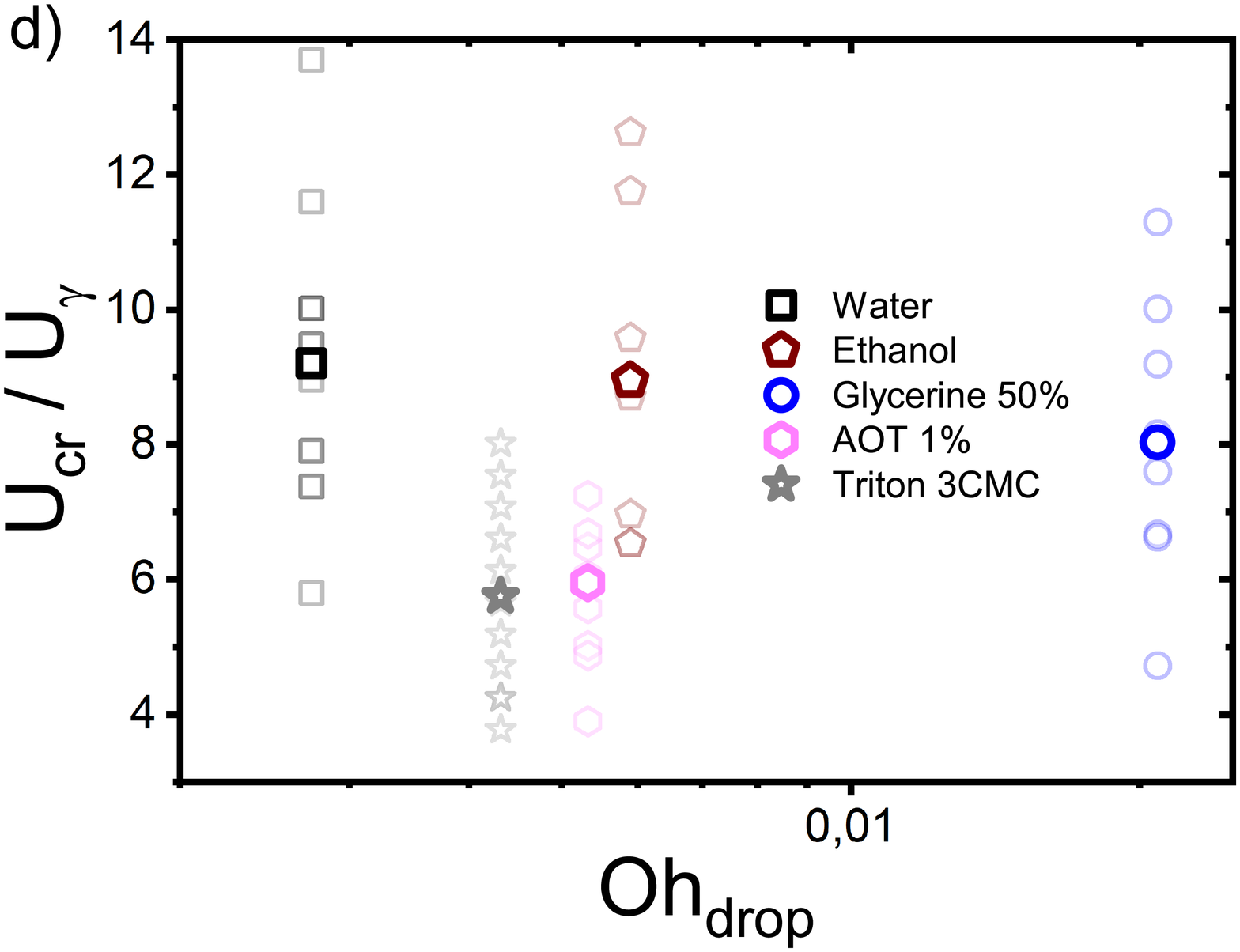}
\caption{Sketch illustrating the definitions of  a) the cavity velocity $U_c$ and b) the velocity of the retracting cavity $U_{cr}$. c) Ratio of the cavity velocity $U_c$ and the impact velocity $U_{jet}$, compared to the expected value 0.5. d) $U_{cr}$ divided by the capillary velocity scale $U_{\gamma} = \sqrt{\gamma_{dyn}/\rho_{drop}D_{drop}}$; average values for each liquid are statistically similar, indicating that the retraction of the cavity is governed by capillary forces. For each liquid, bold coloured symbols represent average values of light symbols.}
\label{Fig8}
\end{figure}

To obtain more insight into the dynamics of the cavity that is created in the droplet, we studied the cavity velocity in the positive $x$ direction, i.e. while the jet is penetrating into the droplet, as sketched in figure \ref{Fig8}a. For each liquid we plot the average ratio of the cavity velocity $U_c$ and the jet velocity $U_{jet}$ (bold symbols), together with the values obtained for each individual experiment (light symbols) as a function of $Oh_{drop}$ in figure \ref{Fig8}c. The measured and averaged values are remarkably close to the value $U_c/U_{jet} = 0.5$, which is to be expected for the impact of a continuous jet on a pool, and is in agreement with previous works \citep{oguz1995air}. The slight deviation observed for the water and the PEO solutions droplets, could be due to water and PEO solution droplets being the largest ones used in the experiments. In that case, the breakup of the jet could influence $U_c$, similarly to a train of droplets impacting a deep pool \citep{bouwhuis2016impact, speirs2018water}. 

In addition to $U_c$, we measured the retraction cavity velocity $U_{cr}$ after the cavity reached its maximum length, as sketched in figure \ref{Fig7}b. In figure \ref{Fig8}d, we show $U_{cr}$  rescaled by the capillary velocity scale $U_{\gamma} = \sqrt{\gamma_{dyn}/\rho_{drop}D_{drop}}$. We observe that the average of data for the different liquids are similar, taking into consideration the data dispersion. The average of ethanol, water and aqueous glycerol mixtures are even statistically indistinguishable, given the error margins of the experiment. The lower average values of $U_{cr}$ for the AOT and Triton solutions can possibly be explained by the Marangoni stresses generated by the flow from areas with low surface tension to those with high surface tension. Indeed, Marangoni stresses have been shown to retard cavity collapse and slow the velocity of Worthington jets \citep{jia2020marangoni, constante-amores_2021}. Therefore, we can assume that $U_{cr} \propto U_{\gamma}$, indicating that the retraction of the cavity is surface tension driven \citep{michon2017jet}. The origin of the dispersion in $U_{cr}$ is associated with the jet tail breakup, where the matryoshka effect or the creation of an antibubble may arise, like in figure \ref{Fig5} \citep{speirs2018entry, song2020criteria}. 

\subsection{Observations after the cavity collapse}
\label{Observations}

After the retraction phase, the cavity collapses and generates a Worthington jet (as e.g., depicted in the last panel of figure \ref{Fig3}). Extensive studies of the length, speed and breakup time of a Worthington jet formed after droplet and solid impact on a liquid pool have been widely reported, and are outside of the scope of this paper \citep{worthington1908study, cheny1996extravagant, gekle2010generation, michon2017jet, che2018impact,mohammad2020experimental, jia2020marangoni, kim2021impact}. Moreover, given the random breakup of the impacting jet in our experiments, the Worthington jets are observed to vary widely in size and shape, even when droplet and impacting jet consist of the same liquids. This is understandable, as it has been shown in the literature that small disturbances in the cavity can have a strong influence on the Worthington jet properties \citep{michon2017jet}. 

\begin{figure}
\centering
\includegraphics[width=13cm]{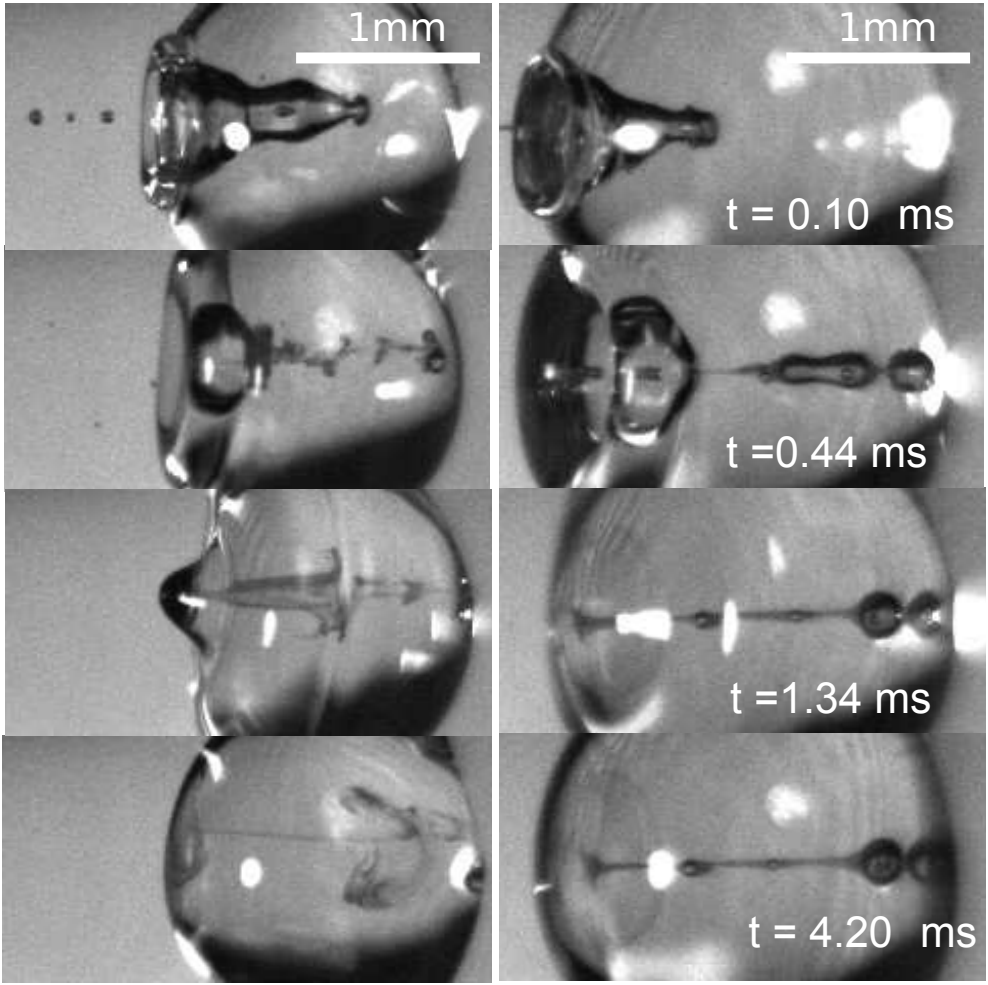}
\caption{Mixing after impact and cavity collapse. Left, jet impacting an AOT 0.1 wt.$\%$ droplet with $We_{jet} = 359$. In the sequence we observe vortical mixing presumably due to the Marangoni flow caused by the difference of surface tension between the jet and the droplet (see movie 6 in the supplementary materials and \cite{jia2020marangoni}). Right, jet impacting a PEO 600k 1 wt.$\%$ with $We_{jet} = 574$. In this sequence we observe little mixing and diffusion due to the viscoelastic properties of the droplet even after 0.42 ms.}
\label{Fig9}
\end{figure}

Lastly, we observe that the mixing and diffusion of the impacting jet into the droplet is governed by the droplet characteristics. Indeed, for a jet impacting a water droplet with AOT 0.1 wt$\%$ below the critical value $U_{jet}^*$ needed for traversing, there is vortical mixing (figure \ref{Fig9} left). Comparable mixing patterns were observed (data not shown) for the rest of the Newtonian liquids containing surfactants, and weaker mixing is seen for water, ethanol and glycerine 25 $\%$ droplets. Similar mixing patterns have been described in the literature, for example, \cite{jia2020marangoni} reported an interfacial Marangoni flow enhancing the mixing of an impacting droplet and a liquid pool with different surface tensions. 

We note that in our experiments the surface tension of the jet is almost always expected to be different from the surface tension of the droplet, and a Marangoni flow could explain this type of mixing. However, a more in depth study is needed to confirm this hypothesis. In contrast, for the viscoelastic liquids with $De \gtrsim 2 \times 10^{-2}$ and the glycerol mixture liquids with $Oh \gtrsim 0.02$, the jet does not mix with the droplet in the timescale of our experiments (figure \ref{Fig9}b). Furthermore, low viscosity ($Oh \lesssim 0.01$ ) and low surface tension liquids reach equilibrium at a later stage than more viscous liquids ($Oh \gtrsim 0.02$) and with higher surface tension. For example, in figure \ref{Fig9} a PEO 600k 1 wt.$\%$ droplet reaches equilibrium $\sim$ 4 times faster than the AOT droplet. This is expected, as surface tension and viscosity have been observed to affect droplet oscillations \citep{kremer2018simultaneous,arcenegui2019simple}.

\section{Conclusions}

We have presented experimental results of liquid water jets impacting on pendant droplets with different liquid properties. We proposed two models to predict a critical jet impact velocity beyond which the jet traverses the droplet. First, we presented a model based on a simple energy balance between the jet kinetic energy and the change in surface tension of the droplet. The second model is based on the comparison between the Young-Laplace and the dynamic pressures of the cavity made by the penetrating jet, and its shape is described by the two-dimensional Rayleigh equation.

Although the critical velocity predicted in both models is of the same order of magnitude, they differ in their scaling relation with $D_{drop}/D_{jet}$. The difference arises from the different description of the cavity geometry and its associated surface energy. In the energy balance model, a cylindrical shape is assumed, contrasting with the more accurate cavity shape described by the two-dimensional Rayleigh equation. Furthermore, we tested the validity of the second model, by fitting our experimental data with equation (\ref{GammaDyn}), showing good agreement when dynamic surface tension effects are considered, see figure \ref{Fig6}. Therefore, for Newtonian droplets the impact process is initially dominated by inertia and their dynamic surface tension is the major opposing force.

In addition, we investigated viscoelastic effects by using water-based polyethylene-oxide solutions of varied concentrations and molecular weight. For $De \lesssim 4 \times 10^{-3}$, the droplets act as if they were Newtonian. In contrast, for $De \gtrsim 2 \times 10^{-2}$, a greater jet impact speed is necessary to traverse the droplet, indicating that when the capillary and relaxation times are comparable, viscoelastic effects can dominate the traversing phenomena.  Moreover, we observed a distinct transition phenomenon from traversing to embedding, which we called pearling and on which the protruing jet is sucked back into the droplet. 

Next, we investigated the advancing and retraction velocities $U_c$ and $U_{cr}$ of the cavity, confirming previous reports that $U_c/U_{jet} \sim 0.5$ for different liquids. Furthermore, we found that $U_{cr}$ is surface tension driven, with the connotation that for droplets containing surfactants $U_{cr}$ is observed to be slower than for the other liquids that were used, which could be explained by Marangoni stresses. 

Our results are relevant for needle-free injections into skin, because of the challenge in quantifying injection processes in real tissue. Given the opacity of skin, our results bridge the gap between the jet penetration of droplets and real tissue. Our findings could also be translated to jet injections in other soft tissues, e.g.~the eye, where controlling the jet velocity, $U_{jet}^*$, would be essential to avoid undesired tissue damage and ensure successful drug delivery.

\section{Acknowledgements}
This research was funded by the European Research Council (ERC) under the European Union Horizon 2020 research and innovation programme (Grant agreement No. 851630). We thank valuable discussions with Loreto Oyarte-G\'alvez, Javier Rodr\'iguez-Rodr\'iguez, \'Alvaro Mar\'in, Ivo Peters and Detlef Lohse. We also thank Ambre Bouillant and Ali Rezaei for their assistance on the shear viscosity measurements. The guidance of James W. Bales and Andrew Davidhazy through Edgerton's Digital Collection was of great value. David Fernandez Rivas, would like to thank Gareth McKinley for his input at the start of the project and for hosting his stay at the HML and the MIT. 

\bibliographystyle{jfm}
\bibliography{jfmbib}

\begin{thebibliography}{52}
\expandafter\ifx\csname natexlab\endcsname\relax\def\natexlab#1{#1}\fi
\def\au#1{#1} \def\ed#1{#1} \def\yr#1{#1}\def\at#1{#1}\def\jt#1{\textit{#1}}
  \def\bt#1{#1}\def\bvol#1{\textbf{#1}} \def\vol#1{#1} \def\pg#1{#1}
  \def\publ#1{#1}\def\arxiv#1{#1}\def\org#1{#1}\def\st#1{\textit{#1}}

\bibitem[Alvarez {\em et~al.\/}(2010)Alvarez, Walker \&
  Anna]{alvarez2010microtensiometer}
{\sc \au{Alvarez, Nicolas~J}, \au{Walker, Lynn~M} \& \au{Anna, Shelley~L}}
  \yr{2010}  \at{A microtensiometer to probe the effect of radius of curvature
  on surfactant transport to a spherical interface}.  \jt{Langmuir}
  \bvol{26}~(16),  \pg{13310--13319}.

\bibitem[Arcenegui-Troya {\em et~al.\/}(2019)Arcenegui-Troya,
  Belman-Mart{\'\i}nez, Castrej{\'o}n-Pita \&
  Castrej{\'o}n-Pita]{arcenegui2019simple}
{\sc \au{Arcenegui-Troya, J}, \au{Belman-Mart{\'\i}nez, A},
  \au{Castrej{\'o}n-Pita, AA} \& \au{Castrej{\'o}n-Pita, JR}} \yr{2019}  \at{A
  simple levitated-drop tensiometer}.  \jt{Review of Scientific Instruments}
  \bvol{90}~(9),  \pg{095109}.

\bibitem[Aristoff \& Bush(2009)]{aristoff2009water}
{\sc \au{Aristoff, Jeffrey~M} \& \au{Bush, John~WM}} \yr{2009}  \at{Water entry
  of small hydrophobic spheres}.  \jt{Journal of Fluid Mechanics}  \bvol{619},
  \pg{45--78}.

\bibitem[Bergmann {\em et~al.\/}(2006)Bergmann, van~der Meer, Stijnman,
  Sandtke, Prosperetti \& Lohse]{bergmann2006giant}
{\sc \au{Bergmann, Raymond}, \au{van~der Meer, Devaraj}, \au{Stijnman, Mark},
  \au{Sandtke, Marijn}, \au{Prosperetti, Andrea} \& \au{Lohse, Detlef}}
  \yr{2006}  \at{Giant bubble pinch-off}.  \jt{Physical Review Letters}
  \bvol{96}~(15),  \pg{154505}.

\bibitem[Bobylev {\em et~al.\/}(2019)Bobylev, Guzanov, Kvon \&
  Kharlamov]{bobylev2019influence}
{\sc \au{Bobylev, AV}, \au{Guzanov, VV}, \au{Kvon, AZ} \& \au{Kharlamov, SM}}
  \yr{2019} Influence of soluble surfactant on wave evolution on falling liquid
  films.  \bt{In {\em Journal of Physics: Conference Series\/}}, ,  \vol{vol.
  1382},  \pg{p. 012073}. IOP Publishing.

\bibitem[Bouwhuis {\em et~al.\/}(2016)Bouwhuis, Huang, Chan, Frommhold, Ohl,
  Lohse, Snoeijer \& van~der Meer]{bouwhuis2016impact}
{\sc \au{Bouwhuis, Wilco}, \au{Huang, Xin}, \au{Chan, Chon~U}, \au{Frommhold,
  Philipp~E}, \au{Ohl, Claus-Dieter}, \au{Lohse, Detlef}, \au{Snoeijer,
  Jacco~H} \& \au{van~der Meer, Devaraj}} \yr{2016}  \at{Impact of a high-speed
  train of microdrops on a liquid pool}.  \jt{Journal of Fluid Mechanics}
  \bvol{792},  \pg{850--868}.

\bibitem[Che \& Matar(2018)]{che2018impact}
{\sc \au{Che, Zhizhao} \& \au{Matar, Omar~K}} \yr{2018}  \at{Impact of droplets
  on immiscible liquid films}.  \jt{Soft Matter}  \bvol{14}~(9),
  \pg{1540--1551}.

\bibitem[Cheny \& Walters(1996)]{cheny1996extravagant}
{\sc \au{Cheny, JM} \& \au{Walters, K}} \yr{1996}  \at{Extravagant viscoelastic
  effects in the worthington jet experiment}.  \jt{Journal of Non-Newtonian
  Fluid Mechanics}  \bvol{67},  \pg{125--135}.

\bibitem[Constante-Amores {\em et~al.\/}(2021)Constante-Amores, Kahouadji,
  Batchvarov, Shin, Chergui, Juric \& Matar]{constante-amores_2021}
{\sc \au{Constante-Amores, C.R.}, \au{Kahouadji, L.}, \au{Batchvarov, A.},
  \au{Shin, S.}, \au{Chergui, J.}, \au{Juric, D.} \& \au{Matar, O.K.}}
  \yr{2021}  \at{Dynamics of a surfactant-laden bubble bursting through an
  interface}.  \jt{Journal of Fluid Mechanics}  \bvol{911},  \pg{A57}.

\bibitem[Crichton {\em et~al.\/}(2011)Crichton, Donose, Chen, Raphael, Huang \&
  Kendall]{crichton2011viscoelastic}
{\sc \au{Crichton, Michael~L}, \au{Donose, Bogdan~C}, \au{Chen, Xianfeng},
  \au{Raphael, Anthony~P}, \au{Huang, Han} \& \au{Kendall, Mark~AF}} \yr{2011}
  \at{The viscoelastic, hyperelastic and scale dependent behaviour of freshly
  excised individual skin layers}.  \jt{Biomaterials}  \bvol{32}~(20),
  \pg{4670--4681}.

\bibitem[Cu {\em et~al.\/}(2019)Cu, Bansal, Mitragotri \&
  Rivas]{cu2019delivery}
{\sc \au{Cu, Katharina}, \au{Bansal, Ruchi}, \au{Mitragotri, Samir} \&
  \au{Rivas, David~Fernandez}} \yr{2019}  \at{Delivery strategies for skin:
  Comparison of nanoliter jets, needles and topical solutions}.  \jt{Annals of
  Biomedical Engineering}  \pg{pp. 1--12}.

\bibitem[Daerr \& Mogne(2016)]{daerr2016pendent_drop}
{\sc \au{Daerr, Adrian} \& \au{Mogne, Adrien}} \yr{2016}  \at{Pendent\_drop: an
  imagej plugin to measure the surface tension from an image of a pendent
  drop}.  \jt{Journal of Open Research Software}  \bvol{4}~(1).

\bibitem[Deng {\em et~al.\/}(2007)Deng, Anilkumar \& Wang]{deng2007role}
{\sc \au{Deng, Q}, \au{Anilkumar, AV} \& \au{Wang, TG}} \yr{2007}  \at{The role
  of viscosity and surface tension in bubble entrapment during drop impact onto
  a deep liquid pool}.  \jt{Journal of Fluid Mechanics}  \bvol{578},
  \pg{119--138}.

\bibitem[Edgerton(1931)]{edgerton1931stroboscopic}
{\sc \au{Edgerton, Harold~E}} \yr{1931}  \at{Stroboscopic moving pictures}.
  \jt{Electrical Engineering}  \bvol{50}~(5),  \pg{327--329}.

\bibitem[Eggers {\em et~al.\/}(2007)Eggers, Fontelos, Leppinen \&
  Snoeijer]{eggers2007theory}
{\sc \au{Eggers, J}, \au{Fontelos, MA}, \au{Leppinen, D} \& \au{Snoeijer, JH}}
  \yr{2007}  \at{Theory of the collapsing axisymmetric cavity}.  \jt{Physical
  Review Letters}  \bvol{98}~(9),  \pg{094502}.

\bibitem[Ferri \& Stebe(2000)]{ferri2000surfactants}
{\sc \au{Ferri, James~K} \& \au{Stebe, Kathleen~J}} \yr{2000}  \at{Which
  surfactants reduce surface tension faster? a scaling argument for
  diffusion-controlled adsorption}.  \jt{Advances in Colloid and Interface
  Science}  \bvol{85}~(1),  \pg{61--97}.

\bibitem[Fung(2013)]{fung2013biomechanics}
{\sc \au{Fung, Yuan-cheng}} \yr{2013} {\em Biomechanics: mechanical properties
  of living tissues\/}.  \publ{Springer Science \& Business Media}.

\bibitem[Gekle \& Gordillo(2010)]{gekle2010generation}
{\sc \au{Gekle, Stephan} \& \au{Gordillo, Jos{\'e}~Manuel}} \yr{2010}
  \at{Generation and breakup of worthington jets after cavity collapse. part 1.
  jet formation}.  \jt{Journal of Fluid Mechanics}  \bvol{663},  \pg{293}.

\bibitem[Graham {\em et~al.\/}(2019)Graham, McConnell, Limbert \&
  Sherratt]{graham2019stiff}
{\sc \au{Graham, Helen~K}, \au{McConnell, James~C}, \au{Limbert, Georges} \&
  \au{Sherratt, Michael~J}} \yr{2019}  \at{How stiff is skin?}
  \jt{Experimental dermatology}  \bvol{28},  \pg{4--9}.

\bibitem[Grotte {\em et~al.\/}(2001)Grotte, Duprat, Loonis \&
  Pi{\'e}tri]{grotte2001mechanical}
{\sc \au{Grotte, Marie}, \au{Duprat, F}, \au{Loonis, Dominique} \&
  \au{Pi{\'e}tri, Eric}} \yr{2001}  \at{Mechanical properties of the skin and
  the flesh of apples}.  \jt{International Journal of Food Properties}
  \bvol{4}~(1),  \pg{149--161}.

\bibitem[Hamilton(1995)]{hamilton1995needle}
{\sc \au{Hamilton, James~G}} \yr{1995}  \at{Needle phobia: a neglected
  diagnosis}.  \jt{Journal of Family Practice}  \bvol{41}~(2),  \pg{169--182}.

\bibitem[Hogan {\em et~al.\/}(2015)Hogan, Taberner, Jones \&
  Hunter]{hogan2015needle}
{\sc \au{Hogan, Nora~C}, \au{Taberner, Andrew~J}, \au{Jones, Lynette~A} \&
  \au{Hunter, Ian~W}} \yr{2015}  \at{Needle-free delivery of macromolecules
  through the skin using controllable jet injectors}.  \jt{Expert Opinion on
  Drug Delivery}  \bvol{12}~(10),  \pg{1637--1648}.

\bibitem[Jia {\em et~al.\/}(2020)Jia, Sun, Zhang, Yin \&
  Wang]{jia2020marangoni}
{\sc \au{Jia, Feifei}, \au{Sun, Kai}, \au{Zhang, Peng}, \au{Yin, Cuicui} \&
  \au{Wang, Tianyou}} \yr{2020}  \at{Marangoni effect on the impact of droplets
  onto a liquid-gas interface}.  \jt{Physical Review Fluids}  \bvol{5}~(7),
  \pg{073605}.

\bibitem[Joodaki \& Panzer(2018)]{joodaki2018skin}
{\sc \au{Joodaki, Hamed} \& \au{Panzer, Matthew~B}} \yr{2018}  \at{Skin
  mechanical properties and modeling: A review}.  \jt{Proceedings of the
  Institution of Mechanical Engineers, Part H: Journal of Engineering in
  Medicine}  \bvol{232}~(4),  \pg{323--343}.

\bibitem[Kiger \& Duncan(2012)]{kiger2012air}
{\sc \au{Kiger, Kenneth~T} \& \au{Duncan, James~H}} \yr{2012}
  \at{Air-entrainment mechanisms in plunging jets and breaking waves}.
  \jt{Annual Review of Fluid Mechanics}  \bvol{44},  \pg{563--596}.

\bibitem[Kim {\em et~al.\/}(2021)Kim, Lee, Bose, Kim \& Lee]{kim2021impact}
{\sc \au{Kim, Dohyung}, \au{Lee, Jinseok}, \au{Bose, Arijit}, \au{Kim, Ildoo}
  \& \au{Lee, Jinkee}} \yr{2021}  \at{The impact of an oil droplet on an oil
  layer on water}.  \jt{Journal of Fluid Mechanics}  \bvol{906}.

\bibitem[Kiyama {\em et~al.\/}(2019{\natexlab{{\em a\/}}})Kiyama, Endo,
  Kawamoto, Katsuta, Oida, Tanaka \& Tagawa]{kiyama2019visualization}
{\sc \au{Kiyama, Akihito}, \au{Endo, Nanami}, \au{Kawamoto, Sennosuke},
  \au{Katsuta, Chihiro}, \au{Oida, Kumiko}, \au{Tanaka, Akane} \& \au{Tagawa,
  Yoshiyuki}} \yr{2019{\natexlab{{\em a\/}}}}  \at{Visualization of penetration
  of a high-speed focused microjet into gel and animal skin}.  \jt{Journal of
  Visualization}  \bvol{22}~(3),  \pg{449--457}.

\bibitem[Kiyama {\em et~al.\/}(2019{\natexlab{{\em b\/}}})Kiyama, Mansoor,
  Speirs, Tagawa \& Truscott]{kiyama2019gelatine}
{\sc \au{Kiyama, Akihito}, \au{Mansoor, Mohammad~M}, \au{Speirs, Nathan~B},
  \au{Tagawa, Yoshiyuki} \& \au{Truscott, Tadd~T}} \yr{2019{\natexlab{{\em
  b\/}}}}  \at{Gelatine cavity dynamics of high-speed sphere impact}.
  \jt{Journal of Fluid Mechanics}  \bvol{880},  \pg{707--722}.

\bibitem[Kremer {\em et~al.\/}(2018)Kremer, Kilzer \&
  Petermann]{kremer2018simultaneous}
{\sc \au{Kremer, J}, \au{Kilzer, A} \& \au{Petermann, M}} \yr{2018}
  \at{Simultaneous measurement of surface tension and viscosity using freely
  decaying oscillations of acoustically levitated droplets}.  \jt{Review of
  Scientific Instruments}  \bvol{89}~(1),  \pg{015109}.

\bibitem[Krizek {\em et~al.\/}(2020)Krizek, De~Goumo{\"e}ns, Delrot \&
  Moser]{krizek2020needle}
{\sc \au{Krizek, Jan}, \au{De~Goumo{\"e}ns, Fr{\'e}d{\'e}ric}, \au{Delrot,
  Paul} \& \au{Moser, Christophe}} \yr{2020}  \at{Needle-free delivery of
  fluids from compact laser-based jet injector}.  \jt{Lab on a Chip}
  \bvol{20}~(20),  \pg{3784--3791}.

\bibitem[Lee {\em et~al.\/}(1997)Lee, Longoria \& Wilson]{lee1997cavity}
{\sc \au{Lee, M}, \au{Longoria, RG} \& \au{Wilson, DE}} \yr{1997}  \at{Cavity
  dynamics in high-speed water entry}.  \jt{Physics of Fluids}  \bvol{9}~(3),
  \pg{540--550}.

\bibitem[Lorenceau {\em et~al.\/}(2004)Lorenceau, Qu{\'e}r{\'e} \&
  Eggers]{lorenceau2004air}
{\sc \au{Lorenceau, {\'E}lise}, \au{Qu{\'e}r{\'e}, David} \& \au{Eggers, Jens}}
  \yr{2004}  \at{Air entrainment by a viscous jet plunging into a bath}.
  \jt{Physical Review Letters}  \bvol{93}~(25),  \pg{254501}.

\bibitem[Mercuri \& Fernandez~Rivas(2021)]{mercuri2021challenges}
{\sc \au{Mercuri, Magal{\'\i}} \& \au{Fernandez~Rivas, David}} \yr{2021}
  \at{Challenges and opportunities for small volumes delivery into the skin}.
  \jt{Biomicrofluidics}  \bvol{15}~(1),  \pg{011301}.

\bibitem[Michon {\em et~al.\/}(2017)Michon, Josserand \&
  S{\'e}on]{michon2017jet}
{\sc \au{Michon, Guy-Jean}, \au{Josserand, Christophe} \& \au{S{\'e}on,
  Thomas}} \yr{2017}  \at{Jet dynamics post drop impact on a deep pool}.
  \jt{Physical Review Fluids}  \bvol{2}~(2),  \pg{023601}.

\bibitem[Mohammad~Karim(2020)]{mohammad2020experimental}
{\sc \au{Mohammad~Karim, Alireza}} \yr{2020}  \at{Experimental dynamics of
  newtonian and non-newtonian droplets impacting liquid surface with different
  rheology}.  \jt{Physics of Fluids}  \bvol{32}~(4),  \pg{043102}.

\bibitem[Mohizin \& Kim(2018)]{mohizin2018current}
{\sc \au{Mohizin, Abdul} \& \au{Kim, Jung~Kyung}} \yr{2018}  \at{Current
  engineering and clinical aspects of needle-free injectors: A review}.
  \jt{Journal of Mechanical Science and Technology}  \bvol{32}~(12),
  \pg{5737--5747}.

\bibitem[Oguz {\em et~al.\/}(1995)Oguz, Prosperetti \& Kolaini]{oguz1995air}
{\sc \au{Oguz, Hasan~N}, \au{Prosperetti, Andrea} \& \au{Kolaini, Ali~R}}
  \yr{1995}  \at{Air entrapment by a falling water mass}.  \jt{Journal of Fluid
  Mechanics}  \bvol{294},  \pg{181--207}.

\bibitem[Oyarte~G{\'a}lvez {\em et~al.\/}(2019)Oyarte~G{\'a}lvez,
  Bri{\'o}~P{\'e}rez \& Fern{\'a}ndez~Rivas]{oyarte2019high}
{\sc \au{Oyarte~G{\'a}lvez, Loreto}, \au{Bri{\'o}~P{\'e}rez, Maria} \&
  \au{Fern{\'a}ndez~Rivas, David}} \yr{2019}  \at{High speed imaging of solid
  needle and liquid micro-jet injections}.  \jt{Journal of Applied Physics}
  \bvol{125}~(14),  \pg{144504}.

\bibitem[Oyarte~G{\'a}lvez {\em et~al.\/}(2020)Oyarte~G{\'a}lvez, Fraters,
  Offerhaus, Versluis, Hunter \& Fern{\'a}ndez~Rivas]{oyarte2020microfluidics}
{\sc \au{Oyarte~G{\'a}lvez, Loreto}, \au{Fraters, Arjan}, \au{Offerhaus,
  Herman~L}, \au{Versluis, Michel}, \au{Hunter, Ian~W} \&
  \au{Fern{\'a}ndez~Rivas, David}} \yr{2020}  \at{Microfluidics control the
  ballistic energy of thermocavitation liquid jets for needle-free injections}.
   \jt{Journal of Applied Physics}  \bvol{127}~(10),  \pg{104901}.

\bibitem[Padilla-Martínez {\em et~al.\/}(2014)Padilla-Martínez,
  Berrospe-Rodriguez, Aguilar, Ramirez-San-Juan \&
  Ramos-Garcia]{padilla2014optic}
{\sc \au{Padilla-Martínez, JP}, \au{Berrospe-Rodriguez, C}, \au{Aguilar, G},
  \au{Ramirez-San-Juan, JC} \& \au{Ramos-Garcia, R}} \yr{2014}  \at{Optic
  cavitation with cw lasers: A review}.  \jt{Physics of Fluids}
  \bvol{26}~(12),  \pg{122007}.

\bibitem[Prausnitz {\em et~al.\/}(2004)Prausnitz, Mitragotri \&
  Langer]{prausnitz2004current}
{\sc \au{Prausnitz, Mark~R}, \au{Mitragotri, Samir} \& \au{Langer, Robert}}
  \yr{2004}  \at{Current status and future potential of transdermal drug
  delivery}.  \jt{Nature Reviews Drug discovery}  \bvol{3}~(2),  \pg{115--124}.

\bibitem[Rastopov \& Sukhodolsky(1991)]{rastopov1991sound}
{\sc \au{Rastopov, Stanislav~F} \& \au{Sukhodolsky, Anatoly~T}} \yr{1991} Sound
  generation by thermocavitation-induced cw laser in solutions.  \bt{In {\em
  Optical Radiation Interaction with Matter\/}}, ,  \vol{vol. 1440},  \pg{pp.
  127--134}. International Society for Optics and Photonics.

\bibitem[Rodr{\'\i}guez {\em et~al.\/}(2017)Rodr{\'\i}guez, Visser,
  Schlautmann, Rivas \& Ramos-Garcia]{rodriguez2017toward}
{\sc \au{Rodr{\'\i}guez, Carla~Berrospe}, \au{Visser, Claas~Willem},
  \au{Schlautmann, Stefan}, \au{Rivas, David~Fernandez} \& \au{Ramos-Garcia,
  Ruben}} \yr{2017}  \at{Toward jet injection by continuous-wave laser
  cavitation}.  \jt{Journal of Biomedical Optics}  \bvol{22}~(10),
  \pg{105003}.

\bibitem[Sokolowski {\em et~al.\/}(2010)Sokolowski, Giovannitti \&
  Boynes]{sokolowski2010needle}
{\sc \au{Sokolowski, Chester~J}, \au{Giovannitti, Joseph~A} \& \au{Boynes,
  Sean~G}} \yr{2010}  \at{Needle phobia: etiology, adverse consequences, and
  patient management}.  \jt{Dental Clinics}  \bvol{54}~(4),  \pg{731--744}.

\bibitem[Song {\em et~al.\/}(2017)Song, Ju, Luo, Han, Dong, Wang, Gu, Zhang,
  Hao \& Jiang]{song2017controlling}
{\sc \au{Song, Meirong}, \au{Ju, Jie}, \au{Luo, Siqi}, \au{Han, Yuchun},
  \au{Dong, Zhichao}, \au{Wang, Yilin}, \au{Gu, Zhen}, \au{Zhang, Lingjuan},
  \au{Hao, Ruiran} \& \au{Jiang, Lei}} \yr{2017}  \at{Controlling liquid splash
  on superhydrophobic surfaces by a vesicle surfactant}.  \jt{Science Advances}
   \bvol{3}~(3),  \pg{e1602188}.

\bibitem[Song {\em et~al.\/}(2020)Song, Zhang \& Wang]{song2020criteria}
{\sc \au{Song, Youngsup}, \au{Zhang, Lenan} \& \au{Wang, Evelyn~N}} \yr{2020}
  \at{Criteria for antibubble formation from drop pairs impinging on a free
  surface}.  \jt{Physical Review Fluids}  \bvol{5}~(12),  \pg{123601}.

\bibitem[Speirs {\em et~al.\/}(2018{\natexlab{{\em a\/}}})Speirs, Mansoor,
  Hurd, Sharker, Robinson, Williams \& Truscott]{speirs2018entry}
{\sc \au{Speirs, NB}, \au{Mansoor, MM}, \au{Hurd, RC}, \au{Sharker, SI},
  \au{Robinson, WG}, \au{Williams, BJ} \& \au{Truscott, Tadd~T}}
  \yr{2018{\natexlab{{\em a\/}}}}  \at{Entry of a sphere into a
  water-surfactant mixture and the effect of a bubble layer}.  \jt{Physical
  Review Fluids}  \bvol{3}~(10),  \pg{104004}.

\bibitem[Speirs {\em et~al.\/}(2018{\natexlab{{\em b\/}}})Speirs, Pan, Belden
  \& Truscott]{speirs2018water}
{\sc \au{Speirs, Nathan~B}, \au{Pan, Zhao}, \au{Belden, Jesse} \& \au{Truscott,
  Tadd~T}} \yr{2018{\natexlab{{\em b\/}}}}  \at{The water entry of
  multi-droplet streams and jets}.  \jt{Journal of Fluid Mechanics}
  \bvol{844},  \pg{1084}.

\bibitem[Truscott {\em et~al.\/}(2014)Truscott, Epps \&
  Belden]{truscott2014water}
{\sc \au{Truscott, Tadd~T}, \au{Epps, Brenden~P} \& \au{Belden, Jesse}}
  \yr{2014}  \at{Water entry of projectiles}.  \jt{Annual Review of Fluid
  Mechanics}  \bvol{46},  \pg{355--378}.

\bibitem[Wang {\em et~al.\/}(2019)Wang, Si, Luo, Dong \&
  Jiang]{wang2019wettability}
{\sc \au{Wang, Ting}, \au{Si, Yifan}, \au{Luo, Siqi}, \au{Dong, Zhichao} \&
  \au{Jiang, Lei}} \yr{2019}  \at{Wettability manipulation of overflow behavior
  via vesicle surfactant for water-proof surface cleaning}.  \jt{Materials
  Horizons}  \bvol{6}~(2),  \pg{294--301}.

\bibitem[Worthington(1908)]{worthington1908study}
{\sc \au{Worthington, Arthur~Mason}} \yr{1908} {\em A study of splashes\/}.
  \publ{Longmans, Green, and Company}.

\bibitem[Zhu {\em et~al.\/}(2000)Zhu, O{\u{g}}uz \&
  Prosperetti]{zhu2000mechanism}
{\sc \au{Zhu, Yonggang}, \au{O{\u{g}}uz, Hasan~N} \& \au{Prosperetti, Andrea}}
  \yr{2000}  \at{On the mechanism of air entrainment by liquid jets at a free
  surface}.  \jt{Journal of Fluid Mechanics}  \bvol{404},  \pg{151--177}.

\end{thebibliography}

\end{document}